\documentclass[pdflatex,sn-basic,Numbered]{sn-jnl}

\usepackage{graphicx}
\usepackage{amsmath,amssymb}
\usepackage{booktabs}

\newcommand{\FormalTitle}{Auditing two-stage MS/MS molecular retrieval on MassSpecGym v1.5: Reranking gains, candidate priors and spectrum dependence}
\newcommand{\FormalShortTitle}{Auditing two-stage MS/MS molecular retrieval}

\newcommand{\FinalSystem}{support-gated feasibility reranker}
\newcommand{\FinalSystemCap}{Support-gated feasibility reranker}

\newcommand{\RowEnsemble}{Stage~1 ensemble (SCCF)}

\newcommand{\RowFinal}{\textbf{\FinalSystemCap{} (final)}}

\begin{document}

\title[\FormalShortTitle]{\FormalTitle}

\author[1]{\fnm{Jiaqi} \sur{Xue}}
\email{aLpkx@outlook.com}

\author[2]{\fnm{Yutong} \sur{Wu}}
\email{3230101243@zju.edu.cn}

\author*[3]{\fnm{Qiang} \sur{Zhang}}
\email{qiang.zhang.cs@zju.edu.cn}

\affil[1]{\orgname{Independent Researcher}}

\affil[2]{\orgdiv{Department of Chemistry},
  \orgname{Zhejiang University},
  \orgaddress{\city{Hangzhou}, \state{Zhejiang} \postcode{310058},
  \country{China}}}

\affil*[3]{\orgdiv{ZJU-UIUC Institute},
  \orgname{Zhejiang University},
  \orgaddress{\city{Haining}, \state{Zhejiang} \postcode{314400},
  \country{China}}}

\abstract{\textbf{Background:} Two-stage retrieval of molecules from tandem
mass spectra (MS/MS) first shortlists mass-compatible candidates and then
reranks them. Benchmark accuracy can reflect candidate priors, replicate
spectra and precursor-mass information as well as fragment evidence. We
audited where the gains of such a system come from on the official
MassSpecGym v1.5 split.

\textbf{Results:} Stage~1 aligned data-safe DreaMS spectrum vectors with
a composite molecular fingerprint. Stage~2 combined a pair reranker,
candidate-relative coverage and a feasibility tree. On 17,556 official-test
spectra, the final system reached 31.05\% R@1, gaining 10.05 points over the
Stage~1 ensemble. A different molecule's spectrum matched on adduct and
precursor mass lowered R@1 by 9.0 points (15.3 with molecules weighted equally).
A training-mean spectrum retained 13.75\% R@1; a candidate-only baseline
reached 7.87\%.
Mass-feasibility features improved test R@1 by 2.52 points
over the top-5 score-only tree, which already included coverage; learned
peak--fragment similarity did not improve on constant coverage.

Post-hoc analyses found higher front-rank accuracy than
an MSAlign ensemble but lower R@20. Rank-1 errors clustered by molecule
(Stage~1 intraclass correlation 0.74). Removing close training neighbours of
200 initially well-covered validation molecules lowered Stage~1 R@20 from
91.0\% to 62.0\% relative to size-matched random removal. Precursor-mass
fusion and molecule-balanced weighting also changed the comparisons.

\textbf{Conclusions:} Mass-feasibility reranking improved front-rank retrieval
and increased query-spectrum dependence. The audit separates these gains
from shortlist limitations and candidate priors; learned structural
discrimination remains unestablished.

\textbf{Scientific contribution:} Within an established two-stage retrieval
framework, this study uses matched controls and spectrum replacements to show
that reranking gains come from integrating candidate-relative mass-feasibility
evidence and increase reliance on the query spectrum, while learned structural
discrimination is not established. A controlled training-coverage intervention
shows that close training neighbours contribute causally to retrieval
reachability for well-covered molecules. The study also shows that per-molecule
weighting and precursor-mass information can change conclusions when methods
are compared.}


\maketitle

\section{Introduction}\label{sec:introduction}

Tandem mass spectrometry (MS/MS) is central to small-molecule annotation, but
mapping a fragment spectrum to a complete molecular structure is
underdetermined. Candidate retrieval makes the problem explicit: for each
query spectrum, a method ranks a protocol-defined library of mass-compatible
structures. MassSpecGym provides standardized spectra, candidate sets,
structure-aware splits, and candidate-level ranking metrics for this setting
\cite{bushuiev2024massspecgym}. Recent systems combine pretrained spectrum or
molecule encoders with multimodal alignment, generative retrieval, and
forward-spectrum models \cite{bushuiev2026dreams,zhang2025glmr,wang2026glacier}.

Classical database-search pipelines already combined molecular-formula or
fingerprint prediction with structure-database ranking, while fragmentation
tools generated candidate-specific substructures or fragmentation hypotheses
\cite{duhrkopSearchingMolecularStructure2015,duhrkopSIRIUS4Rapid2019,
ridderAutomaticCompoundAnnotation2014}. MIST combines chemical-formula-based
spectrum representations with substructure-fingerprint prediction and
contrastive fine-tuning for retrieval
\cite{goldmanAnnotatingMetabolitemass2023}. More recent contrastive systems learn
joint spectrum--molecule spaces directly, including CMSSP, JESTR, CSU-MS2, and
fine-grained alignment in FLARE
\cite{chenCMSSPContrastiveMass2024,kaliaJESTRJointEmbedding2025,
xieCSUMS2ContrastiveLearning2025,chenFLAREFinegrainedLearning2026}.
MVP jointly learns molecular-graph, fingerprint, single-spectrum and
consensus-spectrum views \cite{chenLearningAllViews2026}; its use of consensus
spectra and, in its primary FormSpec setting, ground-truth-formula-based
peak annotations differs from the single-query-spectrum, no-ground-truth-formula
protocol used here. Candidate selection by mass does not itself remove
formula information from a model's spectral inputs.

Two-stage retrieval already appears in FLARE, which reranks candidates from
an embedding search with fine-grained alignment, and in GLMR, which uses
generated structures to rerank pre-retrieved candidates
\cite{chenFLAREFinegrainedLearning2026,zhang2025glmr}. We study
candidate-conditioned mass-feasibility evidence within this division of
labour, with matched controls and audits of candidate priors and spectrum
dependence. A dual encoder is
attractive at library scale because the spectrum is encoded once and molecule
vectors can be precomputed, but a single global vector cannot ask which
candidate-specific fragment explains an observed peak. A second stage can
reserve such candidate-conditioned evidence for a short list of plausible
structures, without simulating spectra for every database molecule. The
division has a direct consequence for evaluation: a reranker can reorder only
the candidates that the retriever places in its shortlist. Rank-1 accuracy
and shortlist recall should therefore be reported separately. Reranking can
improve rank-1 accuracy, but cannot recover a ground-truth molecule absent
from the shortlist.

For the first stage, the molecular representation is a design choice rather
than a drop-in backend. A representation that is useful for property
prediction or generic chemical similarity need not separate the near-mass
candidates that compete in retrieval. De Waele et al.\ showed that
fingerprint-reconstruction fidelity and exact candidate retrieval are
non-equivalent objectives \cite{dewaele2026optimizing}, and MSAlign showed
that contrastive training against each query's own mass-compatible candidates
substantially outperformed generic in-batch negatives
\cite{krzakala2026msalign}. We adopt that candidate-level objective and use a
structurally complementary composite fingerprint (SCCF), which combines
count-based local environments, topological atom pairs, paths and structural
keys, as the Stage~1 molecular target. SCCF was chosen in a historical
screen in which explicit fingerprints outperformed frozen language-model
embeddings, and its Morgan radius was fixed by a matched comparison on
official validation. We report this screen as design context: it does not
isolate the contribution of individual blocks or of block normalisation, and
it does not establish that fingerprints are generally superior to learned or
graph encoders.

Absolute accuracy on such a benchmark is not by itself evidence of MS/MS
identification. \emph{MassSpecGym in the Wild} showed that inconsistent
SMILES canonicalization, candidate ordering and popularity, and component
overlap can create large apparent gains, and introduced MassSpecGym v1.5
together with data-safe components \cite{liu2026massspecgymwild}. We take that
work as our evaluation starting point. Three further questions follow. First,
a model can perform well because candidate lists themselves carry a learnable
prior. Replacing the query spectrum by zero or a training mean, or by the
spectrum of a different molecule with the same adduct and precursor mass,
measures how much of the accuracy depends on the query spectrum; a
candidate-only model that never sees a spectrum measures the prior directly.
Second, MassSpecGym contains unequal numbers of replicate spectra per
compound. Query-weighted metrics answer how often a spectrum is identified,
whereas connectivity-balanced metrics answer how often a compound is, and the
two can rank methods differently. Third, the exact precursor mass still
discriminates among candidates inside the mass window used to build the
lists. Its contribution must therefore be measured when comparing methods
that do and do not use it.

This study audits a two-stage retrieval system under these constraints on the
official MassSpecGym v1.5 split with a data-safe DreaMS spectrum encoder.
Stage~1 is a three-model SCCF ensemble; Stage~2 is a \FinalSystem{} that
combines a pair reranker, a candidate-relative coverage score and a
gradient-boosted tree over mass-feasibility features of candidate fragments
(Fig.~\ref{fig:overview}).
The system, its controls and the primary analysis plan were fixed on official
training and validation data before the primary official-test evaluation. We ask where the system's gains
come from and how much of its accuracy depends on the query spectrum, on
candidate priors, on training coverage and on evaluation choices. The
contributions are:

\begin{enumerate}
\item a frozen official-test evaluation of the final system against the
  Stage~1 ensemble that supplies its shortlist, single-model MSAlign and
  EmbCos, complemented by later aggregation-matched ensembles of both, separating
  front-rank gains from shortlist recall;
\item matched component controls that measure the increment from
  mass-feasibility features beyond score and coverage information,
  together with tests of learned peak--fragment evidence
  against matched controls;
\item spectrum-replacement and candidate-prior controls for every method,
  including full recomputation of the complete system under each replacement;
\item a molecule-level error analysis and a coverage intervention that tests
  the causal contribution of close training neighbours to Stage~1
  reachability; and
\item an analysis of how query weighting, connectivity balancing and a
  precursor-mass prior change the comparisons.
\end{enumerate}

Analyses added after the test evaluation are identified in Methods and the
relevant figure and table captions.
The study does not claim a universally superior molecular representation or
a benchmark free of candidate priors.

\section{Methods}\label{sec:methods}

\subsection{Task, data and evaluation}\label{sec:methods-task}

For query spectrum $x_i$ with its precursor $m/z$ and adduct, the task is to
rank a mass-derived candidate set $\mathcal{C}_i=\{m_{i1},\ldots,m_{in_i}\}$
so that the measured molecule appears as high as possible. We use
MassSpecGym v1.5 \cite{bushuiev2024massspecgym,liu2026massspecgymwild} with
its official split: 194,119 training, 19,429 validation and 17,556 test
spectra. Candidate lists are the released mass-based lists capped at 256
molecules (mean 255.2 on validation). A ground truth that is absent from a
list is never inserted, and candidate recall is reported separately. The
molecular formula of the query is not used as an input or as a filter.

Molecules are identified by RDKit 2025.09.6 canonical isomeric SMILES
\cite{rdkit2025096}. Because
MassSpecGym contains replicate spectra of the same compound, we also group
queries by structural connectivity, recomputed with RDKit as the first block
of the InChIKey rather than taken from stored identifiers. Validation and test
contain 3,184 and 2,997 connectivities, and neither shares a connectivity with
the training partition.

We report R@1, R@5 and R@20, the fraction of queries whose ground truth is
ranked within the top $k$, and the mean reciprocal rank (MRR). Query-weighted
metrics treat each spectrum as one observation; connectivity-balanced metrics
first average within each connectivity and then across connectivities, so
compounds with many replicate spectra do not dominate. Exact score ties are
scored by their expectation under a uniformly random tie order rather than by
the physical order of the candidate file. When a reranker is compared with a
named baseline, a \emph{rescue} is a query moved to rank~1 and a
\emph{damage} a baseline-correct query moved off rank~1; the damage rate is the
fraction of baseline-correct queries damaged. Gap closure is the R@1 gain
divided by the baseline's R@20$-$R@1 difference, i.e.\ the share of the
reachable but misranked queries that a reranker repairs.

Part of the design evidence comes from an earlier formula-group split of the
same data, with a different spectrum checkpoint (Additional file~1,
Section~S6.1). Those results are labelled historical and
are never pooled with official-split results.

\subsection{Stage~1: retrieval with a composite fingerprint target}
\label{sec:methods-stage1}

\begin{figure}[!tp]
\centering
\includegraphics[width=\linewidth]{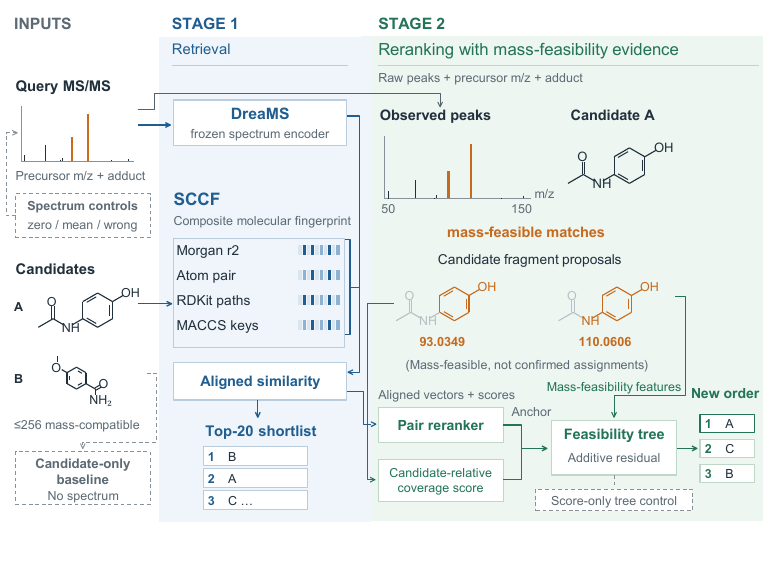}
\caption{Two-stage molecular retrieval and its controls. Stage~1 aligns
frozen data-safe DreaMS spectrum vectors with SCCF, comprising count
Morgan-r2, count atom-pair, RDKit path-count and MACCS-key blocks; the
Stage~1 ensemble supplies the shortlist. Stage~2 combines the pair reranker
and candidate-relative coverage score into an anchor score, then adds a
feasibility-tree residual from score and mass-feasibility features. Only
shortlist order changes. Orange highlights mark mass-feasible proposals,
not confirmed fragment assignments. The support-gated fallback, omitted
from the schematic, uses the top-20 feasibility reranker when supported;
otherwise it checks the original Stage~1 top-5 and, when supported, uses its
own pair reranker, tree and shortlist-relative features; otherwise it uses
top-20 pair + coverage. Support denotes fixed input-applicability conditions,
not confidence. Grey dashed elements are controls: the score-only tree
already includes coverage information, the candidate-only baseline is an
independent spectrum-free control, and spectrum replacement requires
recomputing the entire system.}
\label{fig:overview}
\end{figure}

Figure~\ref{fig:overview} summarises the complete system.

\paragraph{Spectrum encoder.}
Spectra are encoded by DreaMS \cite{bushuiev2026dreams}, a seven-layer
Transformer over a synthetic precursor token and at most the 100 most intense
fragment peaks, with peak $m/z$ and relative intensity as inputs and pairwise
$m/z$ differences in attention. Adduct, instrument and collision energy are
not inputs. We use the data-safe checkpoint released with MassSpecGym v1.5
(\texttt{embedding\_model\_massspecgym.ckpt}), which is matched to the official
split, and take the 1,024-dimensional precursor-token representation before
the released projection head. The encoder is frozen.

\paragraph{Molecule target.}
The structurally complementary composite fingerprint (SCCF) maps each
candidate to four feature blocks implemented in RDKit \cite{rdkit2025096}:
count-based Morgan circular fingerprints \cite{rogers2010ecfp}, topological
atom pairs \cite{carhart1985atompairs}, RDKit path counts and MACCS keys,
\begin{equation}
\begin{aligned}
 f(m)=\operatorname{concat}\big[&
 \operatorname{norm}_2\{\log(1+f_{\mathrm{Morgan},r=2})\},\
 \operatorname{norm}_2\{\log(1+f_{\mathrm{AtomPair}})\},\\
 &\operatorname{norm}_2\{\log(1+f_{\mathrm{path}})\},\
 \operatorname{norm}_2(f_{\mathrm{MACCS}})\big].
\end{aligned}
\end{equation}
Morgan, atom-pair and path-count blocks each have 2,048 hashed count
dimensions and MACCS has 167 keys, giving 6,311 inputs. Morgan and atom-pair
features include chirality; RDKit paths use lengths 1--7 with explicit
hydrogens. Each block is $L_2$-normalised separately so that no block
dominates by its count scale. The blocks describe local atom environments,
pairwise topological distances, paths and structural keys; the name reflects
this intended complementarity and is not a claim of a universally optimal
representation. Morgan radius~2 was fixed by a matched radius comparison on
official validation; the earlier development screen used radius~3.

\paragraph{Alignment and training.}
Two projection heads map the spectrum representation $h_i$ and the molecule
feature to unit vectors,
\begin{equation}
 z_s(x_i)=\operatorname{norm}_2(g_s(h_i)),\qquad
 z_m(m)=\operatorname{norm}_2(g_m(f(m))),
\end{equation}
where $g_s$ is a 1,024$\to$2,048$\to$1,024 and $g_m$ a 6,311$\to$2,048$\to$1,024
multilayer perceptron (GELU, dropout 0.2, layer normalisation). For the ground
truth $m_i^+$ and 127 non-ground-truth molecules $\mathcal{N}_i$ sampled from
the query's own candidate list, training minimises the 128-way candidate
cross-entropy
\begin{equation}
 \mathcal{L}_i=-\log
 \frac{\exp(z_s(x_i)^\top z_m(m_i^+)/\tau)}
 {\exp(z_s(x_i)^\top z_m(m_i^+)/\tau)+
  \sum_{m\in\mathcal{N}_i}\exp(z_s(x_i)^\top z_m(m)/\tau)},
 \qquad \tau=0.07.
\end{equation}
Following MSAlign \cite{krzakala2026msalign}, negatives come from the same
mass-compatible list that defines the retrieval task, so training concentrates
on plausible decoys rather than on unrelated molecules from other batches; we
regard this objective as an established protocol component rather than a
contribution of this study. Negatives whose connectivity occurs in the
validation or test partitions are excluded from sampling. Because this uses
held-out identities (not their spectra), we describe the setting as
transductive with respect to candidate identities. Each model is trained for
one epoch over all training spectra with batch size 128 and AdamW (learning
rate and weight decay $10^{-4}$). Three models are trained with different
random seeds; all candidate features and projected vectors are
precomputed and protected by cache digests.

\paragraph{Stage~1 ensemble and shortlist.}
Each checkpoint $b\in\{1,2,3\}$ scores candidate $j$ by
$\ell_{bj}=z_s^{(b)}(x)^\top z_m^{(b)}(m_j)/\tau$. The Stage~1 ensemble score is
the unweighted mean $s_j=\frac{1}{3}\sum_{b}\ell_{bj}$ over the complete
candidate list; it has no fitted weight or temperature. The natural top-5 or
top-20 of $s$ is the shortlist for Stage~2. Stage~2 reorders candidates only
inside the shortlist, so candidates outside it keep their Stage~1 order and
shortlist recall is fixed.

\subsection{Stage~2: support-gated feasibility reranking}
\label{sec:methods-stage2}

Stage~2 combines three components on the shortlist: a pair reranker that
re-reads the aligned Stage~1 vectors, a coverage score that contrasts
mass-feasible peak support across candidates, and a
gradient-boosted tree that learns a correction from 36 score and
mass-feasibility features. The final system, the \FinalSystem, applies these
components to the top-20 shortlist when its inputs are supported and uses
the support-gated fallback otherwise.

\paragraph{Pair reranker.}
The pair reranker uses the frozen aligned vectors of the first SCCF checkpoint:
the 1,024-dimensional query vector $u$ and candidate vector
$v_j$. A single checkpoint is used to limit computation; it is the first seed
of the prespecified order and was not selected by its validation score. The
candidate input concatenates $[u,\,v_j,\,|u-v_j|,\,u\odot v_j]$ with 11
relative score features computed within the shortlist: for each checkpoint,
the standardised logit and its margin to the shortlist maximum (six features);
across checkpoints, the mean, standard deviation and range of the standardised
logits (three); and the standardised ensemble score and its margin to the
maximum (two). A shared multilayer perceptron
(4,107$\to$256$\to$64$\to$1, GELU, dropout 0.1) adds a residual to the
ensemble score,
\begin{equation}
 p_j=s_j+f_\theta\bigl(u,v_j,\mathbf{q}_j\bigr),
\end{equation}
where $\mathbf{q}_j$ are the score features. The output layer is initialised
at zero, so an untrained reranker reproduces the ensemble order. Training
minimises softmax cross-entropy over the shortlist with the ground truth as
target, using training queries whose ground truth lies in the shortlist and
the same negative exclusion as Stage~1. AdamW (learning rate
$3\times10^{-4}$, weight decay $10^{-4}$, batch 1,024) runs for at most five
epochs. The number of epochs is chosen on a connectivity-disjoint fifth of the
training data as the epoch with the highest R@1 among those with a damage rate
of at most 8\%, where zero epochs (the ensemble itself) is admissible; the
reranker is then refitted on all eligible training queries. Three reranker
seeds are averaged. Candidate position, candidate identifier and
ground-truth formula are never inputs.

\paragraph{Mass-feasibility edges and the role of adduct.}
For precursor $m_0$, observed peak masses $x_p$ and nonnegative intensities
$I_p$, let $M_j$ be candidate $j$'s neutral monoisotopic mass and $f_{jr}$ the
neutral mass of its $r$th fragment proposal, obtained by deterministic
atom-mask enumeration (one non-ring single-bond cut or two cuts in one ring;
at most 64 proposals). For singly charged adduct $a$, the ion offset is
$\Delta_a=m(\mathrm H)-m_e$ for $[\mathrm{M+H}]^+$ and
$m(\mathrm{Na})-m_e$ for $[\mathrm{M+Na}]^+$, with isotope masses from RDKit.
For hydrogen adjustments $h\in\{-2,\ldots,2\}$ the proposed masses are
\begin{align}
 d_{jrh}&=f_{jr}+\Delta_a+h\,m(\mathrm H),\nonumber\\
 \ell_{jrh}&=M_j-f_{jr}+h\,m(\mathrm H).
\end{align}
A peak--proposal edge is feasible when either the direct-ion comparison
($x_p$ versus $d_{jrh}$) or the neutral-loss comparison ($m_0-x_p$ versus
$\ell_{jrh}$) has an absolute error below 20~ppm for some $h$, the ppm
denominator being the proposed mass. Routes and hydrogen adjustments are
combined by logical OR; their identities and the continuous errors are not
features. A feasible edge is a mass-based possibility, not a verified
fragment assignment. Adduct therefore enters only through these mass
calculations and the support rule below; it is not a learned metadata
feature of SCCF, the pair reranker or the tree. Instrument type and collision
energy are not used, and queries with missing values remain eligible.

\paragraph{Coverage score and anchor score.}
Let $e_{jp}$ count feasible proposals of candidate $j$ at peak $p$,
$h_{jp}=\mathbf{1}[e_{jp}>0]$, $w_p=\sqrt{I_p}$ and $W=\sum_p w_p$. With
per-peak support $u_{jp}=\kappa h_{jp}$, where
$\kappa=\tau_c\log(1+\exp(1/\tau_c))$ and $\tau_c=0.1$, the candidate-relative
coverage score is
\begin{equation}
 c_j=\frac{\sum_p w_p\,[u_{jp}-\operatorname{median}_{k}u_{kp}]_+}
 {\max(W,10^{-8})},
\end{equation}
with the lower median for an even shortlist size. Peaks explained by most
shortlisted candidates therefore contribute little, whereas peaks explained by
a minority contribute contrast. With the within-shortlist standardisation
$z(v)_j=(v_j-\bar v)/\max(\sigma_v,0.02)$, the coverage contribution
$0.25\,z(c)_j$ is added to the pair-reranker score to form the anchor score
\begin{equation}
 a_j=p_j+0.25\,z(c)_j .
\end{equation}
The tree described below learns a correction to $a_j$; it does not replace
Stage~1 and is not a fragmentation-probability model.

\paragraph{Score and mass-feasibility features.}
Each shortlisted candidate receives 36 features. The first ten summarise two
score channels, the Stage~1 ensemble score $s$ and
$g_j=s_j+0.25\,z(c)_j$, by
\begin{equation}
 v_j-\bar v,\quad z(v)_j,\quad v_j-\max_k v_k,\quad
 \frac{\exp(v_j)}{\sum_k\exp(v_k)},\quad \sigma_v,
 \qquad v\in\{s,g\}.
\end{equation}
The channel $g$ is a feature only; it differs from the anchor $a$, which is
built on the pair-reranker score. The remaining 26 mass-feasibility features
are 13 deterministic measurements (Table~\ref{tab:lightgbm-feasibility-features})
and their differences from the shortlist mean. Candidate-exclusive support
asks whether other shortlisted candidates can explain the same peak, whereas
ambiguity counts how many proposals of one candidate explain it.

\begin{table}[t]
\caption{The 13 mass-feasibility measurements, grouped by the aspect of
candidate support they describe. Each also contributes its value minus the
mean over shortlisted candidates, giving 26 features.
$P$ is the number of retained observed peaks, $n_j=\sum_p h_{jp}$ and
$b_{jp}=h_{jp}\mathbf{1}[\sum_k h_{kp}=1]$. Count denominators are floored at
1 and intensity denominators at $10^{-12}$; ratios with a zero unfloored
denominator therefore evaluate to zero. The four mass-region measurements
may overlap.}
\label{tab:lightgbm-feasibility-features}
\begingroup
\small
\renewcommand{\arraystretch}{1.15}
\begin{tabular}{@{}p{0.38\linewidth}p{0.56\linewidth}@{}}
\toprule
Measurement & Definition \\
\midrule
\multicolumn{2}{@{}l}{\emph{Peak matching}}\\
Matched peak count & $n_j$ \\
Matched peak fraction & $n_j/P$ \\
\addlinespace[4pt]
\multicolumn{2}{@{}l}{\emph{Fragment feasibility}}\\
Log feasible-edge count & $\log(1+\sum_p e_{jp})$ \\
Proposals per matched peak & $\sum_p e_{jp}/n_j$ \\
Unambiguous matched fraction & $\sum_p\mathbf{1}[e_{jp}=1]/n_j$ \\
\addlinespace[4pt]
\multicolumn{2}{@{}l}{\emph{Intensity and exclusivity}}\\
Intensity coverage & $\sum_p w_p h_{jp}/W$ \\
Exclusive intensity coverage & $\sum_p w_p b_{jp}/W$ \\
Exclusive peak fraction & $\sum_p b_{jp}/P$ \\
\addlinespace[4pt]
\multicolumn{2}{@{}l}{\emph{Mass distribution}}\\
Mean matched peak mass & $\sum_p x_p h_{jp}/n_j$ \\
Low-$m/z$ intensity coverage & $\sum_p w_p h_{jp}\mathbf{1}[x_p/m_0<1/3]/W$ \\
Middle-$m/z$ intensity coverage & $\sum_p w_p h_{jp}\mathbf{1}[1/3\leq x_p/m_0<2/3]/W$ \\
High-$m/z$ intensity coverage & $\sum_p w_p h_{jp}\mathbf{1}[x_p/m_0\geq2/3]/W$ \\
Near-precursor intensity coverage & $\sum_p w_p h_{jp}\mathbf{1}[x_p/m_0\geq0.9]/W$ \\
\bottomrule
\end{tabular}
\endgroup
\end{table}

\paragraph{Feasibility tree.}
We use LightGBM 4.6.0 \cite{ke2017lightgbm} with the LambdaRank objective
\cite{burges2010lambdamart}, one group per query and binary relevance (ground
truth 1, decoys 0). The anchor $a_j$ is supplied as the initial score, so the
trees learn a residual. Each of three seeds fits 50 trees (depth~4, at most 15
leaves, minimum leaf size 50, learning rate 0.03, L2 penalty 10, feature
fraction 0.8, LambdaRank truncation 5); there is no tree-count search. The
final score is
\begin{equation}
 r_j=a_j+\frac{1}{3}\sum_{b=1}^{3}F_b(\mathbf{x}_j),
\end{equation}
and sorting $r_j$ changes the order only inside the shortlist. A tree trained
on in-sample upstream scores would learn from rankings that are better than
those seen at inference. The trees are therefore fitted on out-of-fold
upstream outputs: 8,000 training connectivities (one spectrum each) are split
into two folds, each scored by SCCF models and pair rerankers retrained without
that fold, and 1,024 further training connectivities, scored by a third set of
models, serve as outer calibration (Additional file~1,
Section~S6.3). Only supported fit queries whose ground truth lies
in the shortlist train the trees; calibration and evaluation retain all
queries. At inference the trees are applied unchanged to the full-training
Stage~1 ensemble and pair reranker.

\paragraph{Top-5 and top-20 rerankers.}
The \emph{top-5 feasibility reranker} applies pair reranker, coverage
contribution and tree to the top-5 shortlist. The \emph{top-20 feasibility
reranker} uses the same recipe and hyperparameters on the top-20 shortlist,
with its own retrained pair rerankers and trees and with every mean,
dispersion, softmax denominator, exclusive-support count and coverage median
recomputed over the 20 candidates. No shortlist size, feature or tree
hyperparameter was searched.

\paragraph{Support and the support-gated fallback.}
A shortlist is \emph{supported} when the query adduct is $[\mathrm{M+H}]^+$ or
$[\mathrm{M+Na}]^+$, its peak list is valid, and every shortlisted candidate
is connected, non-isotopic and has at most 60 heavy atoms. Support is a fixed
input-applicability condition, not a model confidence and not a count of
matched peaks. For an unsupported shortlist the tree residual is zero and the
anchor order is kept. The \FinalSystem{} proceeds as follows:
(i) if the top-20 shortlist is supported, use the top-20 feasibility reranker;
(ii) otherwise, if the original Stage~1 top-5 is supported, use the top-5
feasibility reranker, with its own pair reranker, tree and top-5 features
rather than a truncation of the top-20 result;
(iii) otherwise, keep top-20 pair + coverage.
Support depends only on candidate structures, adduct and spectrum validity;
no ground-truth information or fitted threshold enters the decision.

\paragraph{Matched controls.}
Four fixed controls isolate the components on the same shortlist: the pair
reranker alone; pair + coverage ($a_j$ without a tree); a coverage$\times$2
control that doubles the coverage contribution to $0.50\,z(c)_j$, testing
whether a larger coverage weight suffices; and a score-only tree with the same
anchor, capacity and training data but only the ten score-derived features,
i.e.\ without the additional 26 mass-feasibility features. Because $g$ contains
the coverage score, the score-only tree already receives coverage
information; the comparison with the feasibility tree tests the detailed
measurements. These controls are matched comparisons, not steps of a
cumulative chain.

\subsection{Comparison methods}\label{sec:methods-external}

We selected MSAlign and EmbCos as released candidate-level alignment
references that can rank mass-compatible candidates without the query's
ground-truth molecular formula. They cover two input regimes: pretrained
spectrum/molecule embeddings and binned spectra with explicit fingerprints.
This choice supports comparison under our input contract, rather than an
exhaustive ranking of retrieval methods. Additional file~1, Table~S16 and
Section~S6.2 distinguish the inputs, formula requirements and reproduction
scope of these references and selected related methods.

We adapted both methods to the same official partitions, identities,
candidate lists and negative exclusion. The single-model MSAlign
v2 configuration \cite{krzakala2026msalign} aligns the data-safe DreaMS
embedding, together with learned collision-energy and adduct encodings, with
frozen MolDeBERTa molecule vectors. EmbCos, distributed with the MSAlign code,
aligns binned spectra with binary Morgan fingerprints and uses no
metadata. Both keep their released architectures, objectives and training
budgets, use the same three training seeds as SCCF and select one checkpoint
per seed by validation R@1. We report their seed means; published Gaussian precursor-mass
scores and multi-model score fusion are not part of these rows.
To match the aggregation of our Stage~1, we additionally form the
same unweighted mean-logit ensemble of the three MSAlign and of the three
EmbCos seeds; these rows were added after the official-test evaluation.
These are whole-method comparisons: encoders, metadata, budgets and aggregation
differ and are reported explicitly. Attribution of the reranking gain relies
primarily on the matched within-system controls
(Section~\ref{sec:results-ablation}), which hold the shortlist and comparator
information fixed.

\subsection{Spectrum-replacement and candidate-prior controls}
\label{sec:methods-controls}

To measure how much ranking depends on the query spectrum, each method is
re-evaluated with the spectrum replaced while candidates, caches and ranking
code stay fixed. The \emph{zero spectrum} sets the spectrum representation to
zero (and, for Stage~2, supplies an empty peak list). The
\emph{training-mean spectrum} replaces it by the mean over all training
spectra. The \emph{matched wrong-spectrum} control reassigns spectra
one-to-one between queries of different connectivity that share adduct and
exact precursor mass, under three fixed assignments; real and wrong spectra
are compared on the same eligible queries. Because a donor of different
connectivity can still be a close isomer, the validation wrong-spectrum effect
of Stage~1 is additionally stratified by recipient training proximity and by
donor--target SCCF similarity, using the first two prespecified SCCF
checkpoints; outcomes are averaged over checkpoints and assignments before
connectivity-level aggregation. For the complete system, every
replacement recomputes Stage~1 ranking, shortlist, support, routing and all
Stage~2 scores. Two references contain no spectrum at all: a
\emph{candidate-only baseline}, logistic and gradient-boosting rankers trained
on official training data with training-split occurrence and 24 RDKit and
physical descriptors of each candidate, and \emph{random ranking}.

A positive real-minus-control difference shows that a method uses the
spectrum. It does not show that the absolute accuracy is free of candidate
priors: performance that persists without a spectrum is information carried
by the candidate lists themselves. A separate feature-channel control applies
donor spectra only to the spectrum-derived Stage~2 features while keeping the
real pair and anchor scores, to ask whether the feasibility features
themselves use the query spectrum.

\subsection{Molecule-level error structure and training proximity}
\label{sec:methods-molecule}

Rank-1 correctness was analysed at the molecule level. For connectivities
with at least four spectra, ICC(1) from a one-way random-effects analysis of
variance \cite{shrout1979icc} quantifies how much of the variation in
correctness lies between molecules rather than between replicate spectra of
the same molecule. Molecules are classified as all-wrong, all-right or mixed
across their spectra. The within-molecule oracle gives every spectrum of a
molecule the correct outcome if at least one of its spectra is ranked
correctly; it bounds what perfect selection among existing replicate outcomes
could achieve. Because a lower overall accuracy alone increases the
share of all-wrong molecules, each statistic is compared with a null
distribution from 1,000 permutations of correctness across queries with group
sizes fixed.

Training proximity is the cosine similarity between a molecule and its
nearest training ground truth in the raw SCCF feature space (block-wise then
global $L_2$ normalisation, no learned projection); negatives seen during
training are not counted. On the historical split, the causal role of training
coverage was tested by retraining Stage~1 on nested subsets with narrow or
diverse molecule selection at matched row counts, and by expanding the
training set to all available molecules (Additional file~1,
Section~S6.1). On the official split, an intervention on validation
selected 200 molecules whose nearest training ground truth had proximity of at
least 0.90 and retrained Stage~1 with the frozen protocol using two paired
seeds per arm: one arm excluded every training molecule within cosine 0.80
of them, and the other excluded the same number of spectra drawn from
molecules below 0.80 to every
validation molecule. Molecules whose proximity was unchanged by both removals
served as controls, and the primary endpoint was the difference-in-differences
of R@20 between targeted and random removal, selected versus control
molecules.

\subsection{Learned peak--fragment alignment}\label{sec:methods-local}

We also tested whether learned peak--fragment similarities add information
beyond mass feasibility. Frozen DreaMS peak tokens were aligned with a graph
neural network encoding of candidate fragments, trained on MAGMa annotations
\cite{ridderAutomaticCompoundAnnotation2014} of training spectra with
unmatched peaks masked rather than treated as negatives. The resulting local
score was added to the top-5 pair-reranker score and compared with a matched
coverage-only control that replaces every learned similarity of a feasible
peak--fragment edge by a constant. The architecture, proposal audit and pilot
cohorts are given in Additional file~1, Section~S6.4.

\paragraph{Illustrative validation cases.}
For the illustrations we used only official validation, with eligibility
rules fixed before screening: $[\mathrm{M+H}]^+$ queries on the supported
top-20 branch, ground truth in the shortlist, unique first-ranked candidates
before and after reranking, and two displayed structures of distinct
connectivity with at most 30 heavy atoms each. We sampled one query uniformly
from each of two query-ID-sorted sets, using a single random generator with
a fixed seed: a rescue of the ground truth over the original Stage~1 winner,
and a final-system error with a same-formula winner sharing the ground
truth's identical nonempty mass-feasible peak set. Matching was recomputed
with the frozen proposal code and coverage over all 20 candidates. Up to two
hypotheses per molecule were displayed in decreasing matched-peak intensity.
ChemDraw depictions highlight the retained proposal atom sets. These cases
illustrate system behaviour rather than estimate performance or establish
fragmentation mechanisms; test outputs were not used to choose them.

\subsection{Model development, validation and the official test}
\label{sec:methods-selection}

All fitting and selection used the official training and validation
partitions. Validation was used sequentially across development steps, so
validation results are development estimates rather than independent tests.
Each component had a prespecified validation gate. For the top-20 feasibility
reranker the gate required a positive lower confidence bound on the R@1 gain
over the top-5 feasibility reranker, a nonnegative MRR change and a damage
rate of at most 8\%. The original top-20 configuration met the R@1 and MRR
conditions but failed the damage criterion (8.35\%). We retained it by author
judgement, balancing its R@1 and MRR gains against the additional damage.
After inspecting validation outcomes by support stratum, we specified the
support-gated fallback without further fitting or threshold search. The
resulting final system had a validation damage rate of 4.99\% against the
same top-5 baseline (Additional file~1, Section~S7.1) and was frozen before
the official test evaluation.

We performed one primary official-test evaluation after freezing the methods,
result rows, controls and analysis plan. It includes all 15 prespecified
real-input rows (Additional file~1, Section~S3) and the spectrum-replacement
controls of the final system; no checkpoint, seed, coefficient, support rule
or reported row was chosen from test outcomes. Subsequent analyses and
frozen-model scoring comprised aggregation-matched ensembles and additional
retriever controls, MCES and descriptive strata, molecule-level analyses and
precursor-mass fusion; the coverage intervention and illustrative cases used
validation only. These post-hoc analyses, marked $^\dagger$ in the main-text
tables and figures, did not alter the frozen system
(Additional file~1, Sections~S4, S6.6 and S7).

\subsection{Statistical analysis}\label{sec:methods-stats}

Methods are compared on identical queries and candidate lists. Uncertainty is
estimated with 10,000 paired bootstrap resamples of whole connectivities,
reported as percentile 95\% intervals for query-weighted and
connectivity-balanced estimates. Intervals condition on the fitted models and
do not include training or validation-selection uncertainty. The six
prespecified primary test contrasts are reported without multiplicity
adjustment; differences whose interval includes zero are described as
inconclusive. Single-seed results are labelled as such; where an analysis
uses one checkpoint, it is the first prespecified seed, chosen to limit
computation rather than by its validation score. Seed means and standard
deviations of individual models are reported separately from model
ensembles. Exact seed values are listed in Additional file~1, Section~S6.6.
Evaluation records retain query identifiers, candidate counts,
ground-truth ranks and input and checkpoint checksums. Resource accounting
is described in Additional file~1, Section~S6.7.

\subsection{Use of large language models}\label{sec:methods-ai}

ChatGPT (OpenAI) and Claude (Anthropic) assisted with manuscript preparation
and software implementation and debugging. The authors reviewed and
validated the assisted content and code; all reported experimental results
were obtained by executing the research pipeline on public datasets.

\section{Results}\label{sec:results}

All values are percentages (MRR multiplied by 100) and differences are
percentage points with paired 95\% connectivity-cluster bootstrap intervals.
Unless stated otherwise, results are from the single official-test evaluation
(17,556 spectra, 2,997 connectivities, 100\% candidate recall).
A dagger ($^\dagger$) marks analyses added after that evaluation; validation
results and every omitted row are given in Additional file~1.

\subsection{The final system on the official test}\label{sec:results-test}

\begin{figure}[t]
\centering
\includegraphics[width=\textwidth]{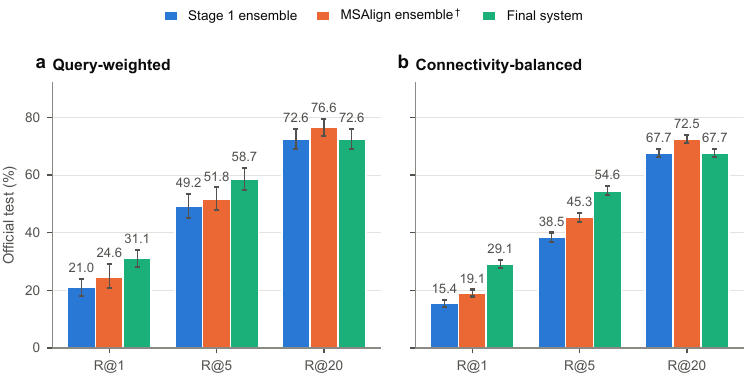}
\caption{Front-rank gains within a fixed shortlist on the official test.
R@1, R@5 and R@20 of the Stage~1 ensemble, the MSAlign ensemble and the final
system, query-weighted (a) and connectivity-balanced (b). Error bars are 95\%
connectivity-cluster bootstrap intervals of each value. The final system
reorders only the Stage~1 shortlist, so its R@20 equals that of the Stage~1
ensemble. $^\dagger$Additional MSAlign ensemble with three-seed mean-logit
aggregation.}
\label{fig:reranking}
\end{figure}

The \FinalSystem{} reached 31.05\% R@1, 58.66\% R@5, 72.62\% R@20 and 43.79\% MRR
(Table~\ref{tab:results-main}). It improved R@1 over the Stage~1 ensemble that
supplies its shortlist by 10.05 points [8.65, 11.51], closing 19.5\% of that
ensemble's R@20$-$R@1 gap, and connectivity-balanced R@1 by 13.61 points
[12.53, 14.69]. Because Stage~2 only reorders the shortlist, R@20 is unchanged
(Fig.~\ref{fig:reranking}).

\begin{table}[t]
\caption{Official-test results. Single-model rows average the metrics of three
independently trained seeds; ensemble rows average the three seeds' logits.
The audited final system is named in bold. Bold/underlined values mark the
highest/second-highest point estimates in each metric column.
Bal.~R@1 is connectivity-balanced R@1. $^\dagger$Ensembles added after the
primary evaluation from frozen per-seed scores.}
\label{tab:results-main}
\centering\small
\setlength{\tabcolsep}{4pt}
\begin{tabular}{@{}lrrrrr@{}}
\toprule
Method & R@1 & R@5 & R@20 & MRR & Bal.~R@1 \\
\midrule
\multicolumn{6}{@{}l}{\emph{Single model (seed mean)}}\\
SCCF & 19.10 & 45.01 & 69.23 & 31.88 & 13.37 \\
MSAlign & 23.88 & 50.18 & \underline{74.56} & 36.53 & 17.75 \\
EmbCos & 15.59 & 32.19 & 54.10 & 24.46 & 13.11 \\
\multicolumn{6}{@{}l}{\emph{Three-seed mean-logit ensemble}}\\
SCCF (Stage~1 ensemble) & 21.01 & 49.20 & 72.62 & 34.51 & 15.45 \\
MSAlign$^\dagger$ & \underline{24.64} & \underline{51.77} & \textbf{76.61} & \underline{37.68} & \underline{19.06} \\
EmbCos$^\dagger$ & 18.40 & 36.28 & 57.67 & 27.68 & 15.21 \\
\midrule
\RowFinal & \textbf{31.05} & \textbf{58.66} & 72.62 & \textbf{43.79} & \textbf{29.05} \\
\bottomrule
\end{tabular}
\end{table}

Against the ensemble of MSAlign, the strongest external retriever
in Table~\ref{tab:results-main}, the final system gained 6.41 R@1 points
[1.16, 10.53], 6.89 R@5 points [4.27, 9.55] and 6.11 MRR points [2.54, 9.09],
but had 3.98 points lower R@20 [$-6.84$, $-1.77$]. Its front-rank advantage
thus coexists with a less complete Stage~1 shortlist. This is a whole-method
comparison: encoders, metadata and training budgets differ. Full seed-mean
contrasts, supplementary ensemble comparisons and the
structural-distance analysis are given in Additional file~1, Sections~S3
and~S4.

\subsection{Stage~2 components and the support-gated fallback}
\label{sec:results-ablation}

Table~\ref{tab:results-ablation} decomposes the final system on the same
test queries. Within the top-5 shortlist, the pair reranker added 1.99 R@1
points to the Stage~1 ensemble and the coverage contribution a further 0.88.
Doubling the coverage weight added 0.97 more, whereas the feasibility tree
reached 29.29\%, 2.52 points above the score-only tree that saw the same scores
and coverage but not the 26 mass-feasibility features. On validation this
feasibility-over-score-only gain was 2.59 points [2.13, 3.08] for the top-20
reranker. Detailed mass-feasibility features thus add information beyond
the score-only tree, which already includes coverage; the coverage$\times$2
control separately tests a larger coverage weight.

\begin{table}[t]
\caption{Component analysis on the official test. All rows rerank the same
Stage~1 ensemble shortlist and share R@20 = 72.62\%. Controls (coverage$\times$2,
score-only tree) are matched comparisons rather than steps of the final
system. Bal.~R@1 is connectivity-balanced. Bold/underlined values mark the
highest/second-highest point estimates in each metric column.}
\label{tab:results-ablation}
\centering\small
\setlength{\tabcolsep}{4pt}
\begin{tabular}{@{}lrrrr@{}}
\toprule
Configuration & R@1 & R@5 & MRR & Bal.~R@1 \\
\midrule
\RowEnsemble & 21.01 & 49.20 & 34.51 & 15.45 \\
\multicolumn{5}{@{}l}{\emph{Top-5 shortlist}}\\
\quad Pair reranker & 23.00 & 49.20 & 35.97 & 16.95 \\
\quad Pair + coverage & 23.88 & 49.20 & 36.74 & 18.09 \\
\quad Coverage$\times$2 control & 24.85 & 49.20 & 37.47 & 19.47 \\
\quad Score-only tree & 26.77 & 49.20 & 38.57 & 21.81 \\
\quad Feasibility reranker & 29.29 & 49.20 & 40.27 & 25.11 \\
\multicolumn{5}{@{}l}{\emph{Top-20 shortlist}}\\
\quad Pair reranker & 22.84 & 50.38 & 36.02 & 16.94 \\
\quad Pair + coverage & 24.19 & 52.92 & 37.71 & 18.86 \\
\quad Coverage$\times$2 control & 25.75 & 55.38 & 39.41 & 21.28 \\
\quad Score-only tree & 27.86 & 56.45 & 41.02 & 24.52 \\
\quad Feasibility reranker & \underline{30.37} & \textbf{58.75} & \underline{43.34} & \underline{28.06} \\
\midrule
\RowFinal & \textbf{31.05} & \underline{58.66} & \textbf{43.79} & \textbf{29.05} \\
\bottomrule
\end{tabular}
\end{table}

Widening the shortlist from 5 to 20 candidates raised R@5 from 49.20\% to
58.75\% and R@1 by 1.08 points, but it also changed which queries were
reranked: relative to the top-5 feasibility reranker, the top-20 version
moved 524 of 5,142 correct predictions off rank~1 (a 10.19\% damage rate)
while rescuing 713. On validation, 190 of the
472 damaged queries belonged to the 2,476 queries whose 20-candidate
shortlist failed the support conditions, so that no tree could be applied,
while their top-5 shortlist met them. The support-gated fallback routes these
queries to the top-5 feasibility reranker, reducing validation damage to
282 queries (4.99\% of the same 5,655 baseline-correct queries).
On test, 13,417 queries used the top-20 branch,
2,396 the top-5 branch and 1,743 retained pair + coverage. Relative to the
top-20 feasibility reranker, the fallback improved R@1 by 0.69 points
[0.45, 0.95] and MRR by 0.45 [0.29, 0.64], and reduced the damage rate relative
to the top-5 reranker from 10.19\% to 6.87\%. This was obtained at a small R@5
cost: $-0.09$ points [$-0.29$, 0.10] query-weighted and $-0.44$
[$-0.72$, $-0.18$] connectivity-balanced.

\subsection{Spectrum dependence and candidate priors}
\label{sec:results-controls}

\begin{figure}[t]
\centering
\includegraphics[width=\textwidth]{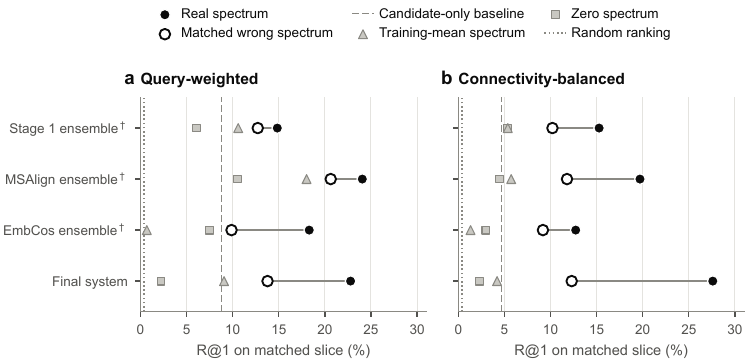}
\caption{Spectrum dependence and candidate priors on the matched
wrong-spectrum test slice (1,887 queries, 301 connectivities), query-weighted
(a) and connectivity-balanced (b) R@1. The matched wrong spectrum comes from a
different connectivity with the same adduct and precursor mass (mean of three
donor assignments). Vertical lines give the spectrum-free candidate-only
baseline and random ranking on the same slice. Additional file~1,
Table~S8 gives uplift ranges and marks intervals excluding zero.
$^\dagger$Additional retriever controls.}
\label{fig:spectrum-controls}
\end{figure}

Every method retained substantial accuracy without the query spectrum
(Fig.~\ref{fig:spectrum-controls}; Additional file~1, Section~S3.3).
On the full test set, the final system
fell from 31.05\% to 2.79\% R@1 with a zero spectrum and to 13.75\% with the
training-mean spectrum. A candidate-only baseline that never sees a spectrum
reached 7.87\%, against 0.40\% for random ranking. Part of the nominal
accuracy is therefore carried by the candidate lists themselves.

The matched wrong-spectrum control isolates query-specific information
(Fig.~\ref{fig:spectrum-controls}): a
spectrum from a different molecule with the same adduct and precursor mass
leaves the candidate list and metadata unchanged. Here the final system
showed the largest and most consistent dependence on the real spectrum,
losing about 9 R@1 points (15 connectivity-balanced) with every interval
above zero. Single-model and ensemble retrievers lost less, and for the
Stage~1 ensemble and the EmbCos ensemble the query-weighted intervals
included zero. Stage~2 thus increases, rather than dilutes, reliance on the
query spectrum. A feature-channel control on validation, which gave donor
spectra only to the mass-feasibility features while keeping the real pair and
anchor scores, reduced R@1 by 3.29--3.43 points with positive intervals; the
features therefore use the query spectrum, although a negligible benefit from
wrong spectra could not be established within the prespecified 0.25-point
margin. Wrong spectra did retain much of the accuracy, especially when donor
and recipient were structurally similar (Additional file~1, Section~S7.4),
consistent with close isomers sharing fragment evidence.

\subsection{Errors cluster by molecule, and training coverage contributes causally}
\label{sec:results-molecule}

\begin{figure}[t]
\centering
\includegraphics[width=\textwidth]{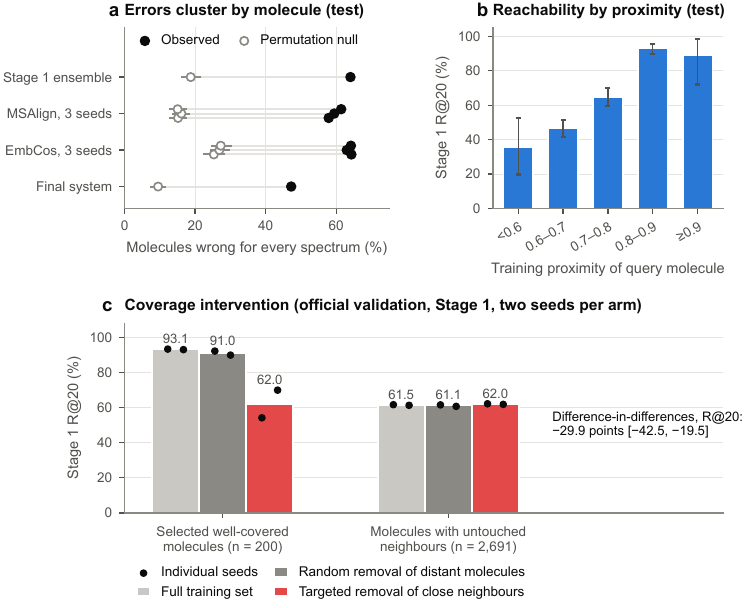}
\caption{Molecule-level errors and training coverage$^\dagger$.
(a)~Share of the 762 test connectivities with at least four spectra for which
every spectrum was wrong at rank~1, observed and under 1,000 permutations of
correctness across queries with group sizes fixed (open circle, mean; line,
95\% range of the permutation distribution). MSAlign and EmbCos show their
three single-model seeds. (b)~Query-weighted Stage~1 ensemble R@20 on the
official test by training proximity of the query molecule, with 95\%
connectivity-cluster intervals; the bins contain 29, 843, 1,459, 562 and 132
connectivities. This association is observational. (c)~Coverage
intervention on official validation: query-weighted Stage~1 R@20 after
retraining without the close positive-training neighbours of 200 initially
well-covered molecules (targeted) or without the same number of spectra from
distant molecules (random), with the existing full-training models for
reference. Bars are two-seed means and dots individual seeds. The
difference-in-differences compares the targeted-minus-random change in the
selected molecules with that in 2,691 molecules whose neighbours were
untouched. The causal interpretation is limited to initially well-covered
molecules and to Stage~1.}
\label{fig:molecule-coverage}
\end{figure}

Rank-1 correctness was largely a property of the molecule rather than of the
individual spectrum (Table~\ref{tab:results-icc}, Fig.~\ref{fig:molecule-coverage}a). Among the 762 test
connectivities with at least four spectra (68\% of queries), the Stage~1
ensemble had ICC(1) 0.74, compared with approximately 0 under permutation, and
64\% of these molecules were wrong for every spectrum, compared with 19\% expected
from its overall accuracy. The same structure appeared for MSAlign and, more
weakly, EmbCos, and on validation (Stage~1 ensemble ICC 0.80). It reproduces
the earlier formula-group result (ICC 0.76). Stage~2 weakened but did not
remove this clustering: the final system lowered the all-wrong share to 47\%,
turning many all-wrong molecules into mixed ones, and the within-molecule
oracle shows that 21.6 further R@1 points would be available if every molecule
were scored by its best spectrum.

\begin{table}[t]
\caption{Molecule-level clustering of rank-1 correctness on the official
test$^\dagger$. ICC and class shares use the 762 connectivities with at least four
spectra; null shares come from 1,000 permutations of correctness across
queries with group sizes fixed. Oracle gain is the R@1 increase if every
spectrum took its molecule's best outcome. MSAlign and EmbCos give the range
over three single-model seeds.}
\label{tab:results-icc}
\centering\footnotesize
\setlength{\tabcolsep}{2pt}
\begin{tabular}{@{}lrrrrr@{}}
\toprule
Method & ICC(1) & All wrong & Null all wrong & Mixed & Oracle gain \\
\midrule
SCCF (Stage~1 ensemble) & 0.74 & 64.0 & 18.8 & 24.8 & 12.5 \\
MSAlign (seeds) & 0.70--0.73 & 57.9--61.4 & 15.0--16.1 & 28.6--33.6 & 14.8--18.8 \\
EmbCos (seeds) & 0.48--0.52 & 63.0--64.3 & 25.3--27.3 & 32.7--34.1 & 20.0--22.0 \\
\RowFinal & 0.64 & 47.2 & 9.5 & 38.1 & 21.6 \\
\bottomrule
\end{tabular}
\end{table}

Training proximity was associated with reachability
(Fig.~\ref{fig:molecule-coverage}b): Stage~1 R@20 rose from
46.5\% for test queries whose molecule had proximity 0.60--0.70 to its nearest
training ground truth to 93.1\% for 0.80--0.90, with a connectivity-level
Spearman correlation of 0.38 [0.35, 0.42]. Proximity accounted for only part of
the molecule-level clustering, however: removing proximity-bin means reduced
the between-molecule variance by 9--21\% across methods on test (7--12\% on
validation).

A coverage intervention tested whether this association is causal
(Fig.~\ref{fig:molecule-coverage}c). On
validation, 200 molecules whose nearest training neighbours had cosine
similarity of at least 0.90 were selected. We retrained Stage~1 with two seeds
per arm (four fits), removing either all training molecules within 0.80 of
them (17,504 spectra) or the same number of spectra drawn from distant
molecules. Compared with random removal, targeted removal lowered their
median proximity by 0.15, R@20 from 91.0\% to 62.0\% and R@1 from
29.6\% to 4.9\%, whereas 2,691 molecules whose neighbours were untouched changed
by less than one point. The difference-in-differences for R@20 was $-29.9$
points [$-42.5$, $-19.5$] query-weighted and $-25.6$ [$-31.0$, $-20.2$]
connectivity-balanced, negative for each seed separately ($-38.6$ and $-21.1$
query-weighted). Close training neighbours therefore causally contribute to
Stage~1 reachability on the official split, consistent with the matched
coverage experiments on the historical split (Additional file~1,
Section~S7.3). The intervention concerns initially well-covered molecules and
Stage~1 only.

\subsection{Weighting by spectrum or by molecule can change the conclusion}
\label{sec:results-weighting}

Because replicate spectra are unevenly distributed across molecules, a
comparison can differ substantially between query-weighted and
connectivity-balanced aggregation (Table~\ref{tab:results-weighting}).
Single-model EmbCos was significantly below SCCF in query-weighted R@1 but
indistinguishable after balancing; the composite SCCF target exceeded a
Morgan-only target on validation by 2.69 query-weighted points but not after
balancing, while its R@5 and R@20 gains persisted; and the candidate-only
baseline lost more than half of its accuracy when each molecule counted once.
The R@5 cost of the fallback was resolved only under balancing.
Query-weighted intervals were also much wider, because a
few heavily replicated molecules dominate them. We therefore report both
weightings throughout and base claims on agreement between them.

\begin{table}[t]
\caption{Selected comparisons under query-weighted and connectivity-balanced
aggregation (R@1 unless stated). Differences with 95\% connectivity-cluster
intervals; absolute values for the candidate-only baseline. Official test
unless marked (validation, three-seed mean). Validation query-weighted SCCF
$-$ Morgan-only differences had query-paired intervals excluding zero.}
\label{tab:results-weighting}
\centering\small
\setlength{\tabcolsep}{3pt}
\begin{tabular}{@{}lll@{}}
\toprule
Comparison & Query-weighted & Balanced \\
\midrule
EmbCos $-$ SCCF (single models) & $-3.51$ [$-6.35$, $-0.33$] & $-0.26$ [$-1.22$, 0.68] \\
SCCF $-$ Morgan-only, validation & $+2.69$ & $-0.18$ [$-0.96$, 0.60] \\
SCCF $-$ Morgan-only R@20, validation & $+1.93$ & $+3.30$ [2.28, 4.31] \\
\textbf{Final} $-$ MSAlign ensemble$^\dagger$ & $+6.41$ [1.16, 10.53] & $+9.99$ [8.59, 11.38] \\
\textbf{Final} $-$ top-20 feasibility, R@5 & $-0.09$ [$-0.29$, 0.10] & $-0.44$ [$-0.72$, $-0.18$] \\
Candidate-only baseline (absolute) & 7.87 & 3.31 \\
\bottomrule
\end{tabular}
\end{table}

\subsection{A precursor-mass prior adds information to every method}
\label{sec:results-mass}

Precursor mass remained informative within the mass-compatible candidate
lists. Adding three Gaussian mass-error scores, with fusion weights
fitted on validation only, raised test R@1 for every
method (Table~\ref{tab:results-mass}). The gain was largest for the Stage~1
retrievers (9--12 points) and smaller for the final system (6.8 points), which
already uses precursor and neutral-loss masses in its feasibility features and
was fused only within its shortlist. The mass score alone ranked the ground truth first for
8.1\% of queries. The precursor mass thus carries substantial
information within the 10-ppm candidate window that is independent of the
fragment spectrum. These supplementary fusion rows are not part of the
frozen system.

\begin{table}[t]
\caption{Precursor-mass prior$^\dagger$ (three Gaussian mass-error scores,
$\sigma=0.01$, 0.1 and 1~ppm) fused with each method's scores, weights fitted
on official validation. Official-test R@1 with 95\% intervals of the gain.
Bold/underlined R@1 values mark the highest/second-highest point estimates
in the Without and With prior columns.}
\label{tab:results-mass}
\centering\small
\setlength{\tabcolsep}{3pt}
\begin{tabular}{@{}lrrl@{}}
\toprule
Method & Without & With prior & Gain \\
\midrule
SCCF (Stage~1 ensemble) & 21.01 & 33.08 & $+12.07$ [10.19, 14.12] \\
MSAlign ensemble & \underline{24.64} & \underline{35.60} & $+10.96$ [9.42, 12.58] \\
EmbCos ensemble & 18.40 & 27.49 & $+9.09$ [6.89, 11.09] \\
\RowFinal & \textbf{31.05} & \textbf{37.88} & $+6.82$ [5.64, 8.07] \\
Mass prior alone & -- & 8.11 & \\
\bottomrule
\end{tabular}
\end{table}

\subsection{Learned local evidence did not add beyond mass feasibility}
\label{sec:results-negative}

Several attempts to replace or extend mass-feasibility evidence with learned
chemistry did not improve on matched controls (Table~\ref{tab:results-negative}).
Learned peak--fragment alignment showed a local semantic signal but did not
outperform a matched coverage-only control under the tested readout: added to
the top-5 pair reranker on validation, it raised R@1 by 0.93 points, compared
with 1.02 for constant similarities and 0.49 for an encoder trained on
shuffled labels, and its rankings agreed with the constant control for 99.1\%
of queries. A limiting factor is the fragment proposal family itself: it
recovered an exact atom mask for only 26.9\% of MAGMa-assigned peaks
(34.8\% of their intensity), and supported candidates had a median of one
mass-feasible peak. Candidate-listwise supervision and fixed feature-group
ablations likewise gave no gain over their matched controls, and among
same-formula queries whose shortlisted candidates had equal coverage the
feasibility tree did not outperform the score-only tree. The gain of the final
system is therefore best described as better integration of mass-feasibility
evidence; learned structural discrimination beyond the matched controls was
not established.

\begin{figure}[!tp]
\centering
\includegraphics[width=\linewidth]{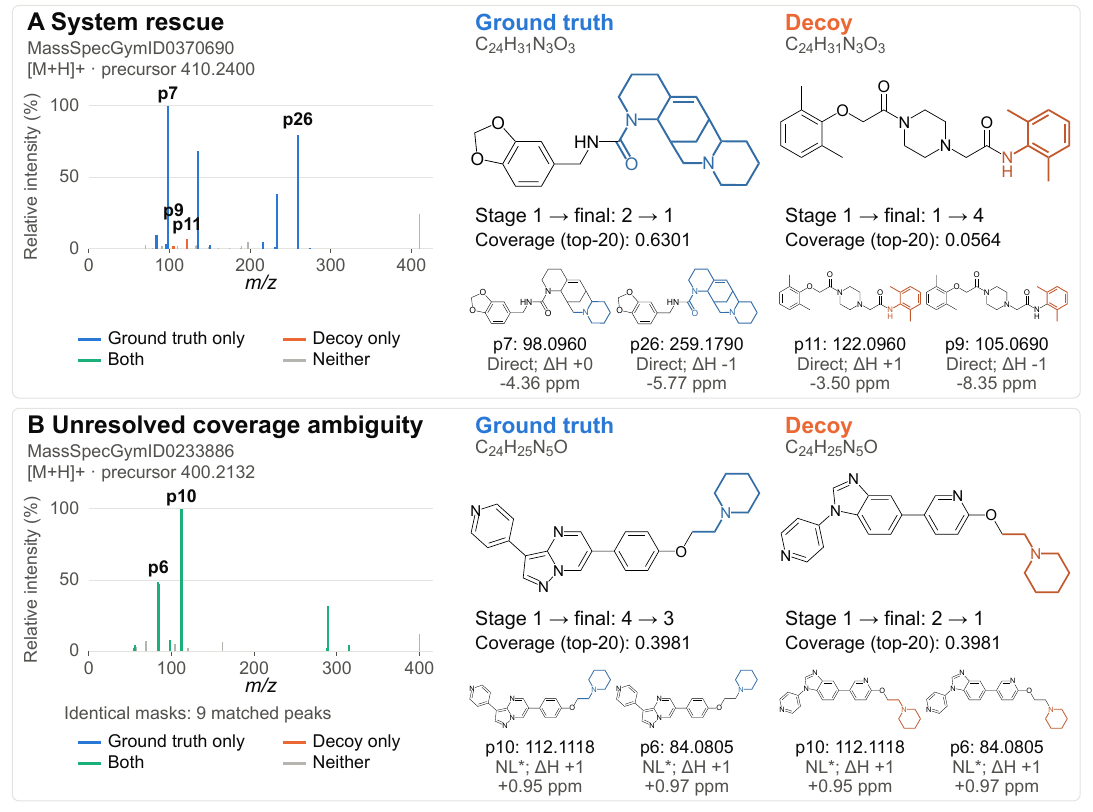}
\caption{Illustrative rescue and unresolved ambiguity on official
validation$^\dagger$. One case per eligible set was sampled with a fixed seed under
the rules in Methods; test outputs were not used. Both use the supported
top-20 branch. (A)~The final system rescues ground truth, which the pair
reranker already ranks first; this is not an isolated coverage effect.
(B)~A same-formula decoy wins despite identical nonempty mass-feasible peak
sets and equal coverage; the failure is specific to this query. Peak colours
distinguish ground-truth-only, decoy-only, shared and unmatched peaks.
Coverage is candidate-relative over all 20 candidates. Up to two hypotheses
per molecule follow decreasing matched-peak intensity; structure highlights
mark their retained atom sets. Direct and NL* denote direct-ion and
neutral-loss mass tests; $\Delta$H is the hydrogen adjustment. These are
mass-feasible hypotheses, not confirmed fragments or losses. Neither means
no match in the frozen proposal family. The cases illustrate behaviour,
not population-level performance.}
\label{fig:validation-cases}
\end{figure}

Two fixed-rule validation examples illustrate how this evidence enters the
ranking (Fig.~\ref{fig:validation-cases}). In the rescue (A), ground truth moves
from Stage~1 rank~2 to final rank~1 and the original winner from rank~1 to~4.
Its coverage is higher, but the pair reranker already places it first, so the
case does not isolate a coverage effect. In the unresolved case (B), ground
truth and a same-formula decoy have equal coverage and identical nonempty
mass-feasible peak sets. Ground truth moves from rank~4 to~3 while the decoy
moves from rank~2 to~1: shared mass-feasibility evidence leaves this query's
structural ambiguity unresolved.

\begin{table}[t]
\caption{Learned or extended local evidence against matched controls
(development evidence; R@1 differences). OOF: out-of-fold training cohort of
2,038 official-training molecules.}
\label{tab:results-negative}
\centering\small
\setlength{\tabcolsep}{3pt}
\begin{tabular}{@{}p{0.44\linewidth}p{0.20\linewidth}p{0.28\linewidth}@{}}
\toprule
Test & Setting & Result \\
\midrule
Learned peak--fragment similarity vs constant coverage & Validation, top-5 & $+0.93$ vs $+1.02$; 99.1\% identical rankings \\
Candidate-listwise + local loss vs local loss only & OOF cohort & $-0.15$ [$-0.39$, 0.05] \\
Candidate descriptors added to feasibility features & OOF cohort & $+0.25$ [$-0.59$, 1.08] \\
Feasibility vs score-only tree, same formula and equal coverage & Validation, 531 queries & $-0.38$ [$-2.13$, 1.38] \\
Four feature-group ablations vs full feature set & Validation & no improvement \\
\bottomrule
\end{tabular}
\end{table}

\subsection{Stage~1 target and computational cost}
\label{sec:results-stage1-cost}

Among molecule targets screened on the historical split, explicit fingerprints
outperformed frozen ChemBERTa embeddings by about 15 R@1 points, and the
composite target improved on binary Morgan fingerprints (Additional file~1,
Section~S7.3). On the official split, the three SCCF seeds reached
20.81$\pm$0.34\% validation R@1, and in a matched radius-3 comparison the
data-safe spectrum checkpoint did not differ significantly from the ordinary
one ($+0.66$ [$-0.96$, 2.27] for the latter). As shown in
Table~\ref{tab:results-weighting}, the composite target mainly deepens the shortlist,
which is what Stage~2 requires.

On validation, the complete system processed one query, from raw spectrum to
ranking, in a median of 35.7~ms (95th percentile 40.5~ms) with cached
molecule projections and fragment proposals, and 347.6~ms when proposals were
first computed. An SCCF model has 19.2M trainable alignment parameters, against
13.8M for MSAlign, which additionally uses a frozen 88.2M-parameter MolDeBERTa
encoder. The SCCF sparse feature arrays occupy 27.8~GiB for 7.2M candidate
molecules. These measurements do not establish an overall efficiency
advantage over the comparison methods (Additional file~1, Section~S7.11).

\section{Discussion}\label{sec:discussion}

This audit suggests three interacting questions for evaluating MS/MS
molecular retrieval: \emph{candidate reachability}, whether the correct
structure enters the shortlist; \emph{spectrum-dependent discrimination},
whether ranking gains depend on query-specific MS/MS evidence; and
\emph{molecule-level generalization}, whether gains extend across molecular
structures rather than being concentrated in heavily replicated compounds.
These questions organise the findings rather than define independent causal
factors. Better front-rank accuracy can coexist with limited shortlist
recall, residual candidate priors or uneven performance across molecules.
Evaluating them together clarifies both where a system improves and what
that improvement establishes about molecular identification.

\paragraph{Reranking and shortlist reachability.}
Shortlist generation and discrimination within that shortlist impose
different constraints. Stage~2 can correct the ordering of reachable
candidates but cannot recover a structure excluded by Stage~1. Its
front-rank gains therefore do not remove the reachability deficit relative
to the MSAlign ensemble (Section~\ref{sec:results-test}). Reporting only R@1
would conceal this distinction and could favour a better reranker while
leaving a less complete shortlist unexamined. Encoders, metadata and training
budgets differ in this whole-method comparison, so the deficit does not
isolate the molecular target. Applying the reranker to stronger retrievers'
shortlists would test whether improved reachability and within-shortlist
ranking can be combined.

\paragraph{Spectrum dependence and candidate priors.}
The spectrum controls separate two questions that absolute accuracy
conflates: whether a method uses the query spectrum, and how much accuracy
persists without it. The matched wrong-spectrum control addresses the first
most specifically because candidate list, adduct and precursor mass stay
fixed. The final system was more dependent on the real spectrum than the
Stage~1 ensemble, and the feature-channel control located part of this
dependence in the mass-feasibility features. Yet training-mean spectra,
matched wrong spectra and a spectrum-free candidate-only model all retained
substantial accuracy. This residual accuracy is consistent with the
candidate-construction and popularity effects documented by
\emph{MassSpecGym in the Wild} \cite{liu2026massspecgymwild}.

Spectrum sensitivity does not, by itself, establish learned structural
discrimination or a fragmentation mechanism. Mass-feasibility matching can
respond to the spectrum without distinguishing structures that explain the
same peaks; the learned peak--fragment controls below make this distinction
concrete. Nor do the replacement controls provide an additive decomposition
of spectral evidence and candidate priors. Close-isomer donors can retain
relevant fragment evidence, and the feature-channel control did not
establish negligible benefit from wrong spectra within its prespecified
margin. Conversely, a zero spectrum is an out-of-distribution input, so poor
performance under that control does not establish freedom from priors.
Negative sampling also excludes held-out candidate identities, a
transductive departure from open-world identification. The supported claim
is spectrum-dependent retrieval in a candidate-prior-affected benchmark.

\paragraph{Molecule-level errors and training coverage.}
Errors clustered by molecule across methods, and Stage~2 weakened this
clustering without removing it. Replicate spectra therefore do not provide
independent evidence of generalization to additional molecular structures.
The within-molecule oracle bounds selection among existing replicate
outcomes, not what a new representation could achieve. Training proximity
provides one link between molecule-level errors and reachability, but its
observational association explains only a minority of the between-molecule
variance. It should not be treated as the dominant cause of all errors.

The coverage intervention strengthens a narrower causal claim: removing
close positive-training neighbours reduced Stage~1 reachability relative to
size-matched removal of distant molecules, with little change in molecules
whose neighbours were untouched. Together with the historical row-matched
diversity experiments, this shows that the availability of related training
structures can affect shortlist inclusion. The intervention selected
initially well-covered validation molecules; it does not show that adding
neighbours helps poorly covered regions, where reachability is lowest.
It also concerns Stage~1 rather than the final system, uses two seeds with
different effect sizes, and leaves negative candidate pools unchanged, so
it does not isolate an MS/MS-specific mechanism. Coverage-stratified
reporting can expose where a system generalizes poorly; expanding training
diversity in those regions remains a separate experimental question.

\paragraph{What the reranker integrates.}
The matched component controls support an increment from detailed
mass-feasibility features beyond the scores and coverage already available
to a score-only tree (Section~\ref{sec:results-ablation}). A feasible edge
records that a proposed fragment or neutral loss could explain a peak by
mass; it is not a verified assignment. This distinction matters when several
candidates offer equally feasible explanations. Learned peak--fragment
similarity did not outperform the constant-similarity coverage control under
the tested readout, despite a local semantic signal. Candidate-listwise
supervision and candidate descriptors likewise gave no gain over matched
controls, and the feasibility tree did not exceed the score-only tree among
same-formula queries with equal coverage. The validation cases illustrate
both a rescue already achieved by the pair reranker and an unresolved
same-formula error with identical feasible peak sets
(Fig.~\ref{fig:validation-cases}). Together, these results distinguish
successful integration of mass-feasibility information from an
as-yet-unestablished gain in learned structural discrimination.

The negative results concern the tested configurations rather than local
evidence in general. One identified constraint is the proposal family,
which recovered an exact atom mask for only about one third of the
MAGMa-assigned intensity, leaving much observed evidence unavailable to a
local scorer. Calibrated forward-spectrum likelihoods, fragment-aware pair
verifiers and empirical spectral-neighbour channels
\cite{wangCFMID40web2022,youngTandemMassspectrum2024,MS2QueryReliablescalable}
remain plausible sources of candidate-discriminative evidence. Any such
channel should be combined with coverage evidence and assessed against
score-only, coverage, same-formula and wrong-spectrum controls. Literature
results of forward models (Additional file~1, Table~S1) use different
protocols and are not compared with our rows.

\paragraph{The Stage~1 target in context.}
SCCF provides the molecular target for the shortlist. Varying this target
while holding the candidate-level objective fixed addresses a question
related to, but distinct from, the finding that fingerprint fidelity and
exact retrieval can favour different solutions \cite{dewaele2026optimizing}.
The historical fingerprint comparison and official-validation gains over
Morgan support SCCF as a shortlist target under this protocol, with R@1
sensitive to the weighting scheme. They do not isolate block normalisation,
show that every block is necessary or establish that fingerprints are
generally superior to learned or graph encoders. A descriptive geometry
diagnostic inspired by representation-landscape analysis
\cite{rottach2025topology} found no resolved advantage of raw SCCF over raw
Morgan in local spectral smoothness (Additional file~1, Section~S7.10), so
that measure does not explain the composite's retrieval behaviour.

\paragraph{Weighting and the benchmark contract.}
Query weighting and connectivity balancing answer different population
questions. The former weights each measured spectrum equally; the latter
averages performance within each connectivity and then weights
connectivities equally. Their disagreements show that a gain across spectra
need not represent a gain across structures. Balancing changed the
interpretation of the composite target's R@1 advantage, candidate-only
accuracy, Stage~1 wrong-spectrum dependence and the fallback's R@5 cost
(Section~\ref{sec:results-weighting}). Collision-energy availability also
tracked replication and difficulty, so its association with accuracy cannot
be assigned to metadata alone (Additional file~1, Section~S7.8).

Together with the candidate cap and positive--decoy priors, these effects
limit interpretation of MassSpecGym v1.5 to ranking within its declared
candidate population. External benchmarks could make query-specific MS/MS
evidence more identifiable by drawing ground truths and property-matched
decoys from one declared library, keeping held-out ground truths out of
training negatives, stratifying near-isomers, and reporting candidate recall
and both weightings. Harmonised resources such as SpectraVerse offer one
foundation for such evaluation \cite{guptaSpectraverse2026}; connectivity
balancing within one benchmark does not itself establish transfer across
libraries or instruments.

\paragraph{Precursor mass as a separate information source.}
The fusion analysis shows that precursor mass remains informative even
within a mass-compatible candidate window. Gaussian mass-error scores can
reweight candidates with different exact masses, but cannot separate
same-formula isomers and do not supply the query formula. Published MSAlign
fusion also uses these scores \cite{krzakala2026msalign}. The final system
already uses precursor and neutral-loss masses, and its fusion was confined
to the shortlist; its smaller gain does not quantify overlap between these
channels. Intervals assess within-method gains rather than superiority
between fused systems. Comparisons should therefore declare precursor-mass
scores separately from MS/MS structural evidence. Our matched wrong-spectrum
controls retain precursor mass while varying fragment evidence.

\paragraph{Limitations.}
First, official validation was used sequentially, including development of
the support-gated fallback, before the final system was frozen for test.
Validation results are therefore development estimates. Second,
test intervals condition on the fitted models, omit training and selection
uncertainty and are not adjusted for multiplicity. Third, the ensemble
comparisons, precursor-mass prior, molecule-level analyses and coverage
intervention were added after the test evaluation, and the intervention used
validation data, one cohort and two seeds. Fourth, external comparisons are
whole-method comparisons, and our reranker was not applied to the external
shortlists. Fifth, candidate priors persist and negative sampling uses
held-out candidate identities, so the results are not a prior-free MS/MS
benchmark. Finally, all evidence comes from one benchmark; transfer to other
libraries, instruments and candidate constructions was not tested.

\section{Conclusions}\label{sec:conclusions}

A higher rank-1 retrieval score does not, by itself, establish better use of
MS/MS evidence. On MassSpecGym v1.5, mass-feasibility reranking improved
front-rank accuracy within a fixed shortlist and increased dependence on the
query spectrum. Yet substantial accuracy persisted with training-mean or
matched wrong spectra, and learned peak--fragment similarity did not
outperform matched coverage controls under the tested readout. The audit
therefore supports effective integration of mass-feasibility information,
while additional discrimination between competing structures through learned
fragmentation evidence remains unestablished.

These findings motivate joint evaluation of candidate reachability,
spectrum-dependent discrimination and molecule-level generalization.
Molecule-level error clustering, the bounded training-coverage intervention
and sensitivity to replicate weighting show why average accuracy across
spectra is insufficient to assess generalization across structures. Causal
support for training coverage remains confined to Stage~1 and initially
well-covered molecules. Future work should combine stronger shortlist
generation with evidence that distinguishes candidates offering equally
feasible explanations of observed peaks, and assess gains across molecules
under explicit candidate-prior and precursor-mass controls.

\backmatter

\bmhead{Supplementary information}

Additional file~1, Supplementary information
(\texttt{supplementary\_information.pdf}), contains all 15 frozen official-test
rows, individual seeds, both weighting schemes, prespecified paired intervals,
and complete-system controls. It also reports the MCES analysis,
including the six comparators restored by frozen-model scoring, descriptive
strata, validation-only latency and negative results. Literature-reported
forward results under unmatched protocols and collision-energy missingness
counts are presented separately.

\bmhead{Abbreviations}

MS/MS, tandem mass spectrometry; SCCF, structurally complementary composite
fingerprint; ICC, intraclass correlation coefficient; MCES, maximum common
edge subgraph; MRR, mean reciprocal rank; R@$k$, recall at rank $k$;
OOF, out of fold; MACCS, Molecular ACCess System structural keys;
ppm, parts per million.

\section*{Declarations}

\subsection*{Ethics approval and consent to participate}
Not applicable.

\subsection*{Consent for publication}
Not applicable.

\subsection*{Availability of data and materials}
MassSpecGym v1.5 is available from the cited dataset snapshot
\cite{massspecgymV15PinnedSnapshot}. Study-specific archived predictions,
statistical summaries, figure source data and selected author-trained
checkpoints are available in the
\href{https://github.com/ChristopherXMag/msms-retrieval-audit/releases/tag/v1.0.0}{version 1.0.0 reproducibility release}.
Author-owned original predictions, statistical summaries and figure source
data are licensed under CC BY 4.0; author-trained model files are licensed
under MIT. Third-party materials retain their own terms. Upstream pretrained
encoders and full embedding caches are not redistributed; their sources and
requirements are documented in the repository.

\subsection*{Code availability}
The source code, configurations and analysis-verification instructions are
available in the
\href{https://github.com/ChristopherXMag/msms-retrieval-audit}{MS/MS retrieval audit repository},
version 1.0.0 (tag \texttt{v1.0.0};
\href{https://github.com/ChristopherXMag/msms-retrieval-audit/commit/e3524f7e62e63b4805d1a577d3076cba27f9ce28}{commit \texttt{e3524f7e62e6}}).
Original software is licensed under MIT. The Python code was developed on
Linux with Python 3.11; dependency specifications and reproduction instructions
are included. Release verification covers archived-prediction analyses and
selected artifact checks, rather than a fresh end-to-end training or inference
run; the tested scope and remaining requirements are documented in the release.

\subsection*{Competing interests}
The authors declare that they have no competing interests.

\subsection*{AI use declaration}
During the preparation of this manuscript, the authors used ChatGPT (OpenAI)
and Claude (Anthropic) to assist with drafting, language refinement, LaTeX
typesetting, and software implementation and debugging. All reported
experimental results were obtained through the execution of the research
pipeline on publicly available datasets, rather than generated directly by
AI tools. All AI-assisted content and code were subject to human review and
validation. The authors take full responsibility for the methodology,
results, and conclusions presented in this work.

\subsection*{Funding}
No specific grant funding was received for this research.

\subsection*{Authors' contributions}
J.X. conceived and designed the study, developed the methodology, implemented
the models, conducted the experiments, analyzed and interpreted the results,
and wrote the original manuscript. Y.W. contributed to discussions of
research ideas and critically reviewed the manuscript. Q.Z. provided
methodological guidance and suggestions throughout the development of the
project. All authors reviewed and approved the final manuscript.

\subsection*{Acknowledgements}
The authors acknowledge the ZJU-UIUC Institute at Zhejiang University for
providing computational resources used in the early stage of this study.

\bibliography{references}

\end{document}


\maketitle
\section{Published forward-model retrieval results}
\label{sec:published-forward-context}

Table~\ref{tab:published-forward-context} provides literature context for
forward-spectrum retrieval. All values are taken from Table~1 of the
\href{https://arxiv.org/html/2606.29161v2}{GLACIER preprint},
version~2 (5~October~2026) \cite{wang2026glacier}; the
ICEBERG~2.0 row is the result reported in that comparison. The source task is
the MassSpecGym mass challenge with up to 256 same-mass candidates and
0.01\,Da spectral bins. These are author-reported results under their source
protocols, not reruns using our frozen candidates, checkpoints or evaluator.

\begin{table}[htbp]
\centering
\caption{Literature-reported forward-model retrieval results
(\%; unmatched protocols). CF denotes contrastive finetuning. Only published
mean recalls are reproduced; the source reports its own 99.9\% bootstrap
intervals. No result from our study is included in this table.}
\label{tab:published-forward-context}
\begin{tabular}{@{}lrrr@{}}
\toprule
Model & R@1 & R@5 & R@20 \\
\midrule
ICEBERG 2.0 & 63.95 & 83.83 & 92.56 \\
GLACIER, without CF & 69.69 & 86.58 & 93.34 \\
GLACIER, with CF & 69.95 & 86.47 & 93.52 \\
\bottomrule
\end{tabular}
\end{table}

\paragraph{Acquisition inputs and comparison scope.}
The released MassSpecGym configurations of ICEBERG and GLACIER enable adduct,
collision-energy (CE) and instrument-type conditioning for spectral prediction
\cite{mspredIcebergMsgConfig,mspredGlacierMsgConfig}.
Our retrieval pipeline does not require CE or instrument-type
annotations at inference. Precursor $m/z$ and adduct information remain inputs
to the mass-feasibility calculations; the pipeline is not metadata-free.
These are differences in input requirements, not evidence that any accuracy
difference is caused by acquisition metadata or that our model is more robust
across instruments. Exact evaluated query membership, preprocessing, candidate
handling, training provenance and missing-condition policies were not matched
to the published forward-model rows. Accordingly, this table does not support
cross-study performance deltas, a common ranking of methods, or a claim of
superiority over forward-spectrum retrieval.

\section{Collision-energy availability in MassSpecGym v1.5}
\label{sec:ce-availability}

In the pinned MassSpecGym v1.5 snapshot,
\textbf{109,358 of 231,104 spectrum rows (47.32\%) lack a scalar CE value}
\cite{bushuiev2024massspecgym,massspecgymV15PinnedSnapshot}.
Table~\ref{tab:ce-availability} separates the whole-dataset count from the
train and validation partitions used in this study. These are spectrum-row
counts, not counts of unique molecules. Missing values retain their missing
status rather than being recoded as zero energy.

\begin{table}[htbp]
\centering
\caption{Missing scalar CE in the pinned MassSpecGym v1.5 metadata.
Percentages are calculated from the reported spectrum-row counts.
No held-out retrieval evaluation was performed for this summary.}
\label{tab:ce-availability}
\begin{tabular}{@{}lrrr@{}}
\toprule
Scope & Spectrum rows & Missing CE & Missing (\%) \\
\midrule
Whole dataset & 231,104 & 109,358 & 47.32 \\
Official train & 194,119 & 92,510 & 47.66 \\
Official validation & 19,429 & 9,454 & 48.66 \\
\bottomrule
\end{tabular}
\end{table}

Our inference interface therefore accommodates the many queries without CE
annotations, without estimating CE for each candidate. A condition-aware
forward model can instead define an explicit imputation or fallback policy;
those choices require separate documentation and are not reconstructed by
Table~\ref{tab:published-forward-context}. Annotation-free inference does not
establish acquisition-independent representations: the observed spectrum
still reflects its acquisition conditions, and missingness can correlate with
data source. The counts above describe the available metadata, not
the performance contribution of CE conditioning.

\paragraph{Dataset version and scope.}
Counts refer to dataset revision
\begin{quote}
\small\texttt{0dc77f670939718f333cd06e8a7b70fdfc5a39a5}.
\end{quote}
This metadata summary uses existing counts without additional access to test
spectra, structure labels or predictions. No model was fitted or evaluated.
Table~\ref{tab:published-forward-context} reports published results only;
we did not perform a matched ICEBERG/GLACIER comparison.

\clearpage
\section{Frozen official-test results}
\label{sec:s7-test}

We evaluated all 17,556 official-test spectra and 2,997 recomputed
connectivities with mass-derived candidate lists capped at 256. Models,
aggregation, the support-gated fallback, all 15 real-input rows and the
analysis plan were fixed before reading test outcomes. The final system is
the \FinalSystem. All retrieval values below are percentages; MRR is
multiplied by 100. Connectivity-balanced R@1 first averages within connectivity
groups.

Validation development of the final system is reported in Section~S7.1.
Candidate-prior effects remain, and the use of held-out
connectivities to exclude training negatives makes training transductive with
respect to candidate identities (main-text Methods).

\subsection{All prespecified real-input rows}

\begin{table}[htbp]
\centering\small\setlength{\tabcolsep}{3pt}
\caption{All 15 frozen real-input rows on official test. Method seed means are
arithmetic means of single-model metrics; the SCCF ensemble instead averages
logits before ranking. Top-5 and top-20 controls retain their own frozen branches.
All rows were fixed before reading test outcomes. The final system is named
in bold.}
\label{tab:s7-all-rows}
\begin{tabular}{@{}lrrrrr@{}}
\toprule Method & R@1 & R@5 & R@20 & MRR & Bal. R@1 \\ \midrule
\RowSCCFMean & 19.1027 & 45.0103 & 69.2318 & 31.8829 & 13.3749 \\
\RowMSAlignMean & 23.8788 & 50.1794 & 74.5595 & 36.5332 & 17.7523 \\
\RowEmbCosMean & 15.5894 & 32.1865 & 54.1045 & 24.4580 & 13.1113 \\
\RowEnsemble & 21.0071 & 49.1969 & 72.6247 & 34.5131 & 15.4480 \\
\RowTopFivePair & 23.0007 & 49.1969 & 72.6247 & 35.9747 & 16.9499 \\
\RowTopFivePairCov & 23.8836 & 49.1969 & 72.6247 & 36.7394 & 18.0950 \\
\RowTopFiveCovDouble & 24.8462 & 49.1969 & 72.6247 & 37.4652 & 19.4669 \\
\RowTopFiveScoreOnly & 26.7658 & 49.1969 & 72.6247 & 38.5722 & 21.8140 \\
\RowTopFiveFeas & 29.2891 & 49.1969 & 72.6247 & 40.2672 & 25.1107 \\
\RowTopTwentyPair & 22.8355 & 50.3759 & 72.6247 & 36.0183 & 16.9435 \\
\RowTopTwentyPairCov & 24.1912 & 52.9164 & 72.6247 & 37.7135 & 18.8583 \\
\RowTopTwentyCovDouble & 25.7519 & 55.3771 & 72.6247 & 39.4082 & 21.2821 \\
\RowTopTwentyScoreOnly & 27.8594 & 56.4536 & 72.6247 & 41.0193 & 24.5164 \\
\RowTopTwentyFeas & 30.3657 & 58.7491 & 72.6247 & 43.3394 & 28.0576 \\
\RowFinal & 31.0549 & 58.6580 & 72.6247 & 43.7938 & 29.0541
\\
\bottomrule
\end{tabular}
\end{table}

\begin{table}[htbp]
\centering\small\setlength{\tabcolsep}{3pt}
\caption{All 15 rows with every metric balanced over recomputed connectivity.}
\label{tab:s7-all-balanced}
\begin{tabular}{@{}lrrrr@{}}
\toprule Method & R@1 & R@5 & R@20 & MRR \\ \midrule
\RowSCCFMean & 13.3749 & 34.7748 & 63.0499 & 24.5128 \\
\RowMSAlignMean & 17.7523 & 43.2904 & 69.5568 & 30.2163 \\
\RowEmbCosMean & 13.1113 & 29.5014 & 51.5794 & 21.9898 \\
\RowEnsemble & 15.4480 & 38.4939 & 67.6850 & 27.2272 \\
\RowTopFivePair & 16.9499 & 38.4939 & 67.6850 & 28.4185 \\
\RowTopFivePairCov & 18.0950 & 38.4939 & 67.6850 & 29.4470 \\
\RowTopFiveCovDouble & 19.4669 & 38.4939 & 67.6850 & 30.5486 \\
\RowTopFiveScoreOnly & 21.8140 & 38.4939 & 67.6850 & 31.8984 \\
\RowTopFiveFeas & 25.1107 & 38.4939 & 67.6850 & 34.1776 \\
\RowTopTwentyPair & 16.9435 & 41.1783 & 67.6850 & 29.0273 \\
\RowTopTwentyPairCov & 18.8583 & 45.3953 & 67.6850 & 31.5244 \\
\RowTopTwentyCovDouble & 21.2821 & 49.2084 & 67.6850 & 34.1956 \\
\RowTopTwentyScoreOnly & 24.5164 & 51.2459 & 67.6850 & 36.8877 \\
\RowTopTwentyFeas & 28.0576 & 55.0032 & 67.6850 & 40.3411 \\
\RowFinal & 29.0541 & 54.5679 & 67.6850 & 40.9329
\\
\bottomrule
\end{tabular}
\end{table}

\subsection{Individual training seeds and their dispersion}

\begin{table}[htbp]
\centering\scriptsize\setlength{\tabcolsep}{2pt}
\caption{Arithmetic mean and sample standard deviation over three independently
trained seeds. This dispersion is distinct from the connectivity-bootstrap
interval and does not describe an ensemble.}
\label{tab:s7-seed-sd}
\begin{tabular}{@{}llrrrr@{}}
\toprule Method & Estimand & R@1 & R@5 & R@20 & MRR \\ \midrule
SCCF & Query & $19.1027\pm 1.4168$ & $45.0103\pm 0.9239$ & $69.2318\pm 0.5959$ & $31.8829\pm 1.0910$ \\
 & Balanced & $13.3749\pm 0.0274$ & $34.7748\pm 0.1166$ & $63.0499\pm 0.6596$ & $24.5128\pm 0.0392$ \\
MSAlign & Query & $23.8788\pm 0.5584$ & $50.1794\pm 0.3560$ & $74.5595\pm 0.0786$ & $36.5332\pm 0.3503$ \\
 & Balanced & $17.7523\pm 0.2299$ & $43.2904\pm 0.3444$ & $69.5568\pm 0.2265$ & $30.2163\pm 0.1912$ \\
EmbCos & Query & $15.5894\pm 0.5814$ & $32.1865\pm 0.4949$ & $54.1045\pm 0.8375$ & $24.4580\pm 0.5600$ \\
 & Balanced & $13.1113\pm 0.0134$ & $29.5014\pm 0.1010$ & $51.5794\pm 0.5330$ & $21.9898\pm 0.1029$
\\
\bottomrule
\end{tabular}
\end{table}

\begin{table}[htbp]
\centering\small\setlength{\tabcolsep}{3pt}
\caption{Every frozen SCCF/MSAlign/EmbCos seed on the same official-test cohort.}
\label{tab:s7-seeds}
\begin{tabular}{@{}lllrrrr@{}}
\toprule Method & Seed & Estimand & R@1 & R@5 & R@20 & MRR \\ \midrule
SCCF & 20260621 & Query & 17.4812 & 43.9508 & 68.6204 & 30.6336 \\
 &  & Balanced & 13.4009 & 34.8705 & 62.6982 & 24.5432 \\
SCCF & 20260622 & Query & 20.1014 & 45.6482 & 69.2641 & 32.6474 \\
 &  & Balanced & 13.3775 & 34.8090 & 62.6407 & 24.4685 \\
SCCF & 20260623 & Query & 19.7254 & 45.4318 & 69.8109 & 32.3678 \\
 &  & Balanced & 13.3463 & 34.6449 & 63.8109 & 24.5266 \\
MSAlign & 20260621 & Query & 24.1456 & 50.5354 & 74.4703 & 36.7450 \\
 &  & Balanced & 18.0124 & 43.6296 & 69.5036 & 30.4284 \\
MSAlign & 20260622 & Query & 23.2371 & 50.1794 & 74.6184 & 36.1288 \\
 &  & Balanced & 17.5764 & 43.3008 & 69.3617 & 30.0571 \\
MSAlign & 20260623 & Query & 24.2538 & 49.8234 & 74.5899 & 36.7257 \\
 &  & Balanced & 17.6681 & 42.9409 & 69.8051 & 30.1633 \\
EmbCos & 20260621 & Query & 16.2546 & 32.7473 & 54.6837 & 25.1043 \\
 &  & Balanced & 13.1059 & 29.4813 & 52.1141 & 22.1084 \\
EmbCos & 20260622 & Query & 15.3356 & 31.8111 & 53.1442 & 24.1163 \\
 &  & Balanced & 13.1013 & 29.4121 & 51.5760 & 21.9233 \\
EmbCos & 20260623 & Query & 15.1781 & 32.0011 & 54.4855 & 24.1536 \\
 &  & Balanced & 13.1266 & 29.6109 & 51.0480 & 21.9378
\\
\bottomrule
\end{tabular}
\end{table}

\subsection{Complete-system spectrum and prior controls}

Each spectrum intervention replaces both the cached DreaMS representation
before learned projection and the raw peak list, then recomputes all upstream
rankings, shortlists, support checks and routing. Zero supplies a zero embedding
and an empty peak list; it is not an encoder output computed from empty peaks.
The training-mean spectrum uses official training data only. The matched
wrong-spectrum slice has 1,887 spectra and 301 connectivities; all three donor
assignments match adduct and precursor mass, differ in connectivity, and form
closed one-to-one permutations without fixed points. These prespecified
test interventions cover the final system; the additional retriever controls
reported in the main text were added after the primary evaluation. Earlier external-method validation
controls retain their original scope. The candidate-only baseline is trained
on training data only and, like the exact random expectation, is evaluated on
the full test cohort. Accuracy retained under these replacements is not an
additive causal decomposition.

\begin{table}[htbp]
\centering\scriptsize\setlength{\tabcolsep}{2pt}
\caption{Complete-system controls for the final system. Matched-slice real and wrong rows
share identical recipients; their absolute levels must not be compared with
full-test real as if the populations were equal.}
\label{tab:s7-controls}
\begin{tabular}{@{}lllrrrr@{}}
\toprule Input / population & Queries & Estimand & R@1 & R@5 & R@20 & MRR \\ \midrule
\RowRealFull & 17556 & Query & 31.0549 & 58.6580 & 72.6247 & 43.7938 \\
 &  & Balanced & 29.0541 & 54.5679 & 67.6850 & 40.9329 \\
\RowZero & 17556 & Query & 2.7911 & 11.0333 & 43.7116 & 9.2731 \\
 &  & Balanced & 1.7439 & 8.1935 & 31.3745 & 6.9965 \\
\RowTrainMean & 17556 & Query & 13.7503 & 31.2258 & 50.2563 & 23.1161 \\
 &  & Balanced & 4.1846 & 12.9558 & 33.1565 & 10.2303 \\
\RowCandidateOnly & 17556 & Query & 7.8690 & 22.3847 & 38.6827 & 15.9352 \\
 &  & Balanced & 3.3081 & 9.9504 & 22.1514 & 7.8836 \\
\RowRandom & 17556 & Query & 0.3978 & 1.9891 & 7.9562 & 2.4255 \\
 &  & Balanced & 0.3967 & 1.9833 & 7.9333 & 2.4194 \\
\RowRealMatched & 1887 & Query & 22.7875 & 60.0424 & 73.5559 & 39.1808 \\
 &  & Balanced & 27.6029 & 58.2320 & 75.9065 & 41.5169 \\
\RowWrongOne & 1887 & Query & 13.7255 & 49.3906 & 65.4478 & 29.6545 \\
 &  & Balanced & 12.2900 & 35.1111 & 57.7868 & 23.5181 \\
\RowWrongTwo & 1887 & Query & 13.8315 & 49.2316 & 65.3948 & 29.7224 \\
 &  & Balanced & 12.3786 & 34.9180 & 57.6760 & 23.5408 \\
\RowWrongThree & 1887 & Query & 13.7785 & 49.3376 & 65.3948 & 29.7187 \\
 &  & Balanced & 12.2263 & 35.0587 & 57.5459 & 23.4692
\\
\bottomrule
\end{tabular}
\end{table}

The matched-slice controls for the Stage~1 ensemble, the additional MSAlign
and EmbCos ensembles, and the final system are compared in
Table~\ref{tab:si-test-retriever-controls}. Figure~3 in the main text shows
the same comparisons under both query-weighted and connectivity-balanced
aggregation.

\begin{table}[htbp]
\caption{Spectrum-replacement controls on the matched wrong-spectrum test
slice (1,887 queries, 301 connectivities), query-weighted R@1. Wrong is the
mean over three donor assignments; the uplift column gives the range of the
three real-minus-wrong differences, and $^\ast$ marks that all three
intervals exclude zero. Candidate-only baseline: 8.80\%; random ranking:
0.39\%. MSAlign keeps the recipient's collision energy and adduct under
replacement. $^\dagger$Additional retriever controls.}
\label{tab:si-test-retriever-controls}
\centering\footnotesize
\setlength{\tabcolsep}{2pt}
\begin{tabular}{@{}lrrrrrr@{}}
\toprule
Method & Real & Wrong & Zero & Train mean & \multicolumn{2}{c}{Real $-$ wrong} \\
\cmidrule(l){6-7}
 & & & & & Query & Balanced \\
\midrule
SCCF (Stage~1 ensemble)$^\dagger$ & 14.84 & 12.72 & 6.04 & 10.60 & 2.07--2.17 & 5.00--5.13$^\ast$ \\
MSAlign ensemble$^\dagger$ & 24.06 & 20.63 & 10.49 & 18.02 & 3.39--3.50$^\ast$ & 7.81--7.98$^\ast$ \\
EmbCos ensemble$^\dagger$ & 18.30 & 9.87 & 7.53 & 0.69 & 8.27--8.52 & 3.53--3.58$^\ast$ \\
\RowFinal & 22.79 & 13.78 & 2.23 & 9.06 & 8.96--9.06$^\ast$ & 15.22--15.38$^\ast$ \\
\bottomrule
\end{tabular}
\end{table}

\clearpage
\subsection{Prespecified paired contrasts}

All six primary contrasts are retained below for both estimands and all four
metrics. Values are percentage-point differences with percentile 95\% intervals
from 10,000 paired whole-connectivity bootstrap draws with a fixed seed. Query
weighting treats spectra as observations; balanced weighting gives each
connectivity equal weight. Intervals condition on fixed fits and omit training
and prior validation-selection uncertainty. The multiplicity family is
reported without correction. The top-20 feasibility reranker and the final
system with its support-gated fallback are reported separately.

\small\setlength{\tabcolsep}{3pt}
\begin{longtable}{@{}p{45mm}llrr@{}}
\caption{All prespecified primary test contrasts.}\label{tab:s7-contrasts}\\
\toprule Contrast & Estimand & Metric & $\Delta$ & 95\% interval \\ \midrule
\endfirsthead
\toprule Contrast & Estimand & Metric & $\Delta$ & 95\% interval \\ \midrule
\endhead
\bottomrule\endfoot
Final minus \RowTopTwentyFeas{} & Query & R@1 & +0.6892 & [+0.4485, +0.9500] \\
 &  & R@5 & -0.0911 & [-0.2878, +0.1019] \\
 &  & R@20 & +0.0000 & [+0.0000, +0.0000] \\
 &  & MRR & +0.4544 & [+0.2891, +0.6449] \\
 & Balanced & R@1 & +0.9965 & [+0.7035, +1.3091] \\
 &  & R@5 & -0.4353 & [-0.7179, -0.1755] \\
 &  & R@20 & +0.0000 & [+0.0000, +0.0000] \\
 &  & MRR & +0.5918 & [+0.4037, +0.7937] \\
Final minus MSAlign (seed mean) & Query & R@1 & +7.1761 & [+2.1056, +11.2160] \\
 &  & R@5 & +8.4786 & [+5.9123, +11.0866] \\
 &  & R@20 & -1.9348 & [-4.8069, +0.2948] \\
 &  & MRR & +7.2606 & [+3.7387, +10.2644] \\
 & Balanced & R@1 & +11.3018 & [+9.9919, +12.6505] \\
 &  & R@5 & +11.2775 & [+9.8358, +12.7245] \\
 &  & R@20 & -1.8718 & [-3.2090, -0.5285] \\
 &  & MRR & +10.7166 & [+9.5942, +11.8742] \\
Final minus \RowTopFiveFeas{} & Query & R@1 & +1.7658 & [+0.7263, +2.6920] \\
 &  & R@5 & +9.4612 & [+8.2553, +10.7079] \\
 &  & R@20 & +0.0000 & [+0.0000, +0.0000] \\
 &  & MRR & +3.5266 & [+2.6291, +4.3374] \\
 & Balanced & R@1 & +3.9434 & [+3.2940, +4.6136] \\
 &  & R@5 & +16.0740 & [+14.9363, +17.2259] \\
 &  & R@20 & +0.0000 & [+0.0000, +0.0000] \\
 &  & MRR & +6.7553 & [+6.1584, +7.3666] \\
Final minus \RowEnsemble{} & Query & R@1 & +10.0478 & [+8.6523, +11.5084] \\
 &  & R@5 & +9.4612 & [+8.2553, +10.7079] \\
 &  & R@20 & +0.0000 & [+0.0000, +0.0000] \\
 &  & MRR & +9.2806 & [+8.1032, +10.5162] \\
 & Balanced & R@1 & +13.6061 & [+12.5315, +14.6898] \\
 &  & R@5 & +16.0740 & [+14.9363, +17.2259] \\
 &  & R@20 & +0.0000 & [+0.0000, +0.0000] \\
 &  & MRR & +13.7057 & [+12.8905, +14.5389] \\
MSAlign minus SCCF (seed means) & Query & R@1 & +4.7761 & [+1.0714, +9.7227] \\
 &  & R@5 & +5.1692 & [+2.7427, +7.5502] \\
 &  & R@20 & +5.3277 & [+3.1707, +8.1270] \\
 &  & MRR & +4.6503 & [+2.1613, +7.5907] \\
 & Balanced & R@1 & +4.3774 & [+3.4382, +5.3241] \\
 &  & R@5 & +8.5156 & [+7.2820, +9.7817] \\
 &  & R@20 & +6.5069 & [+5.3021, +7.6924] \\
 &  & MRR & +5.7035 & [+4.8268, +6.5674] \\
EmbCos minus SCCF (seed means) & Query & R@1 & -3.5133 & [-6.3537, -0.3316] \\
 &  & R@5 & -12.8238 & [-15.9664, -9.7134] \\
 &  & R@20 & -15.1273 & [-18.0451, -12.2892] \\
 &  & MRR & -7.4249 & [-9.5995, -5.2652] \\
 & Balanced & R@1 & -0.2636 & [-1.2179, +0.6767] \\
 &  & R@5 & -5.2734 & [-6.5444, -3.9712] \\
 &  & R@20 & -11.4706 & [-12.8383, -10.1243] \\
 &  & MRR & -2.5229 & [-3.4380, -1.5903]
\\
\end{longtable}
\normalsize

\subsection{Prespecified real-minus-control contrasts}

The same bootstrap recipe applies to every planned control contrast. Full-test
zero, training-mean, candidate-only and random references use all queries; each wrong
spectrum uses the identical matched slice as its real reference. These contrasts
measure spectrum sensitivity and prior effects under their stated interventions,
not a causal decomposition or isolated attribution to mass-feasibility features.

\small\setlength{\tabcolsep}{3pt}
\begin{longtable}{@{}p{45mm}llrr@{}}
\caption{All prespecified real-minus-control test contrasts.}\label{tab:s7-control-contrasts}\\
\toprule Contrast & Estimand & Metric & $\Delta$ & 95\% interval \\ \midrule
\endfirsthead
\toprule Contrast & Estimand & Metric & $\Delta$ & 95\% interval \\ \midrule
\endhead
\bottomrule\endfoot
Real minus zero spectrum & Query & R@1 & +28.2638 & [+25.3530, +31.3117] \\
 &  & R@5 & +47.6247 & [+42.9608, +52.3210] \\
 &  & R@20 & +28.9132 & [+24.7990, +33.4660] \\
 &  & MRR & +34.5207 & [+31.7274, +37.3837] \\
 & Balanced & R@1 & +27.3102 & [+25.8348, +28.7763] \\
 &  & R@5 & +46.3744 & [+44.6925, +48.0663] \\
 &  & R@20 & +36.3105 & [+34.5759, +38.0777] \\
 &  & MRR & +33.9364 & [+32.6461, +35.2424] \\
Real minus training-mean spectrum & Query & R@1 & +17.3046 & [+15.0931, +19.6184] \\
 &  & R@5 & +27.4322 & [+24.0863, +30.7702] \\
 &  & R@20 & +22.3684 & [+19.9192, +25.0803] \\
 &  & MRR & +20.6777 & [+18.4369, +22.9069] \\
 & Balanced & R@1 & +24.8695 & [+23.4450, +26.2688] \\
 &  & R@5 & +41.6121 & [+39.9637, +43.3012] \\
 &  & R@20 & +34.5285 & [+32.8391, +36.2053] \\
 &  & MRR & +30.7026 & [+29.4332, +31.9618] \\
Real minus wrong spectrum, assignment 1 & Query & R@1 & +9.0620 & [+5.8165, +12.9811] \\
 &  & R@5 & +10.6518 & [+7.0200, +16.0365] \\
 &  & R@20 & +8.1081 & [+5.1691, +12.4424] \\
 &  & MRR & +9.5264 & [+6.7456, +13.4056] \\
 & Balanced & R@1 & +15.3129 & [+11.0495, +19.5366] \\
 &  & R@5 & +23.1210 & [+17.9414, +28.2543] \\
 &  & R@20 & +18.1198 & [+13.7872, +22.6345] \\
 &  & MRR & +17.9988 & [+14.1628, +21.8250] \\
Real minus wrong spectrum, assignment 2 & Query & R@1 & +8.9560 & [+5.7425, +12.8371] \\
 &  & R@5 & +10.8108 & [+7.1398, +16.2537] \\
 &  & R@20 & +8.1611 & [+5.1614, +12.5583] \\
 &  & MRR & +9.4584 & [+6.6971, +13.3050] \\
 & Balanced & R@1 & +15.2243 & [+10.9658, +19.4301] \\
 &  & R@5 & +23.3140 & [+18.1058, +28.4567] \\
 &  & R@20 & +18.2305 & [+13.8697, +22.7484] \\
 &  & MRR & +17.9761 & [+14.1489, +21.8088] \\
Real minus wrong spectrum, assignment 3 & Query & R@1 & +9.0090 & [+5.7958, +12.8639] \\
 &  & R@5 & +10.7048 & [+7.0465, +16.1165] \\
 &  & R@20 & +8.1611 & [+5.1293, +12.5579] \\
 &  & MRR & +9.4622 & [+6.7011, +13.3016] \\
 & Balanced & R@1 & +15.3766 & [+11.1060, +19.6162] \\
 &  & R@5 & +23.1733 & [+17.9994, +28.3556] \\
 &  & R@20 & +18.3606 & [+14.0066, +22.8629] \\
 &  & MRR & +18.0477 & [+14.2051, +21.8758] \\
Real minus candidate-only baseline & Query & R@1 & +23.1859 & [+19.8791, +26.6966] \\
 &  & R@5 & +36.2733 & [+31.3874, +40.9522] \\
 &  & R@20 & +33.9420 & [+28.2680, +39.3790] \\
 &  & MRR & +27.8586 & [+24.4690, +31.3998] \\
 & Balanced & R@1 & +25.7460 & [+24.2096, +27.2628] \\
 &  & R@5 & +44.6175 & [+42.8107, +46.4999] \\
 &  & R@20 & +45.5336 & [+43.6159, +47.4642] \\
 &  & MRR & +33.0492 & [+31.6068, +34.4922] \\
Real minus random ranking & Query & R@1 & +30.6571 & [+27.7252, +33.7141] \\
 &  & R@5 & +56.6689 & [+52.7735, +60.5380] \\
 &  & R@20 & +64.6685 & [+61.0579, +67.8477] \\
 &  & MRR & +41.3682 & [+38.5402, +44.1795] \\
 & Balanced & R@1 & +28.6574 & [+27.2168, +30.0974] \\
 &  & R@5 & +52.5846 & [+51.0066, +54.1659] \\
 &  & R@20 & +59.7517 & [+58.2221, +61.2338] \\
 &  & MRR & +38.5135 & [+37.1880, +39.8530]
\\
\end{longtable}
\normalsize

\subsection{Exact ties and numerical reproducibility}

Primary shortlist membership uses a fixed identity-key seed. Uniform tie
expectation is conditional on the actual reranked block and retained tail.
Two additional seeds are sensitivity checks and never a selection rule.
Fractional tied hits yield sharp rescue/damage bounds rather than an assumed
joint tie distribution. Before test inference, the implementation reproduced
all validation outputs bitwise (Section~S6.6).

\clearpage
\section{Post-hoc descriptive analyses and validation latency}
\label{sec:s7-posthoc}

We performed these secondary, descriptive analyses after the prespecified
official-test evaluation. They change no original prediction,
model, route, threshold or primary contrast. Six MCES rows required additional
scoring with frozen comparison models, as described below.

\subsection{Training proximity and natural query strata}

Proximity is cosine similarity to the nearest of 25,046 distinct official-train
ground-truth molecule rows in the frozen sparse radius-2 SCCF space: block-wise
normalization, then global L2 normalization, without learned projection. All
test queries are retained. Bins are descriptive summaries, not deployment gates.
Connectivities may cross bins, adducts or routes. Connectivity balancing is
recomputed within each slice, so balanced slice values cannot be added to
recover the full-test balanced value.

\scriptsize\setlength{\tabcolsep}{3pt}
\begin{longtable}{@{}p{30mm}rrrrrrrr@{}}
\caption{Descriptive strata of test predictions. All retrieval values
are percentages. Stage~1 denotes the Stage~1 ensemble; Final denotes the final
system. Group counts may overlap. No slice-wise significance
or causal claim is made.}\label{tab:s7-posthoc-strata}\\
\toprule
 & & & \multicolumn{2}{c}{Stage~1} & \multicolumn{4}{c}{Final} \\
\cmidrule(lr){4-5}\cmidrule(l){6-9}
Stratum & $n$ & Groups & R@1 & R@20 & R@1 & R@5 & MRR & Bal. R@1 \\ \midrule
\endfirsthead
\toprule
 & & & \multicolumn{2}{c}{Stage~1} & \multicolumn{4}{c}{Final} \\
\cmidrule(lr){4-5}\cmidrule(l){6-9}
Stratum & $n$ & Groups & R@1 & R@20 & R@1 & R@5 & MRR & Bal. R@1 \\ \midrule
\endhead
\multicolumn{9}{l}{\textit{proximity}} \\
$<0.60$ & 78 & 29 & 3.85 & 35.90 & 15.38 & 32.05 & 23.76 & 12.13 \\
{[0.60,0.70)} & 3332 & 843 & 6.48 & 46.55 & 17.77 & 36.13 & 26.74 & 19.38 \\
{[0.70,0.80)} & 6742 & 1459 & 9.82 & 64.79 & 22.31 & 47.78 & 34.50 & 26.84 \\
{[0.80,0.90)} & 5100 & 562 & 38.88 & 93.12 & 46.24 & 78.76 & 60.65 & 44.55 \\
$\geq0.90$ & 2304 & 132 & 35.76 & 89.15 & 42.80 & 79.47 & 59.02 & 53.37 \\
\multicolumn{9}{l}{\textit{adduct}} \\
{[M+H]+} & 14066 & 2898 & 19.44 & 68.70 & 29.89 & 56.29 & 41.97 & 29.84 \\
{[M+Na]+} & 3490 & 489 & 27.34 & 88.42 & 35.73 & 68.19 & 51.15 & 30.47 \\
\multicolumn{9}{l}{\textit{candidate count}} \\
$<128$ & 177 & 18 & 15.82 & 99.44 & 18.08 & 47.46 & 36.64 & 13.05 \\
{[128,256)} & 367 & 51 & 18.26 & 88.28 & 17.71 & 58.86 & 35.41 & 21.68 \\
256 & 17012 & 2930 & 21.12 & 72.01 & 31.48 & 58.77 & 44.05 & 29.28 \\
\multicolumn{9}{l}{\textit{route}} \\
Top-20 branch & 13417 & 2607 & 18.34 & 69.64 & 30.23 & 59.15 & 43.15 & 30.02 \\
Top-5 branch & 2396 & 628 & 31.76 & 83.89 & 37.98 & 61.44 & 50.25 & 29.76 \\
Pair + coverage branch & 1743 & 284 & 26.74 & 80.15 & 27.88 & 51.06 & 39.90 & 18.19
\\
\bottomrule
\end{longtable}
\normalsize

Across 2,997 connectivities, mean proximity is associated with retrieval
accuracy (Table~\ref{tab:s7-posthoc-proximity-correlations}). The association
persists on held-out identities, but the highest query-weighted bins are not
monotonic. Training proximity does not uniquely determine
success. Adduct, pool-size and branch contrasts mix chemical composition,
acquisition and difficulty; no causal adduct effect or ground-truth-dependent routing rule
follows.

\begin{table}[htbp]
\centering\small
\caption{Connectivity-level Spearman associations of mean training
proximity with retrieval accuracy on the official test.}
\label{tab:s7-posthoc-proximity-correlations}
\begin{tabular}{@{}lr@{}}
\toprule Method and metric & Spearman correlation \\ \midrule
Stage~1 ensemble R@20 & 0.3849 \\
Stage~1 ensemble R@1 & 0.3291 \\
Final system R@1 & 0.2532 \\
\bottomrule
\end{tabular}
\end{table}

\subsection{Batch-one complete-system timing on official validation}

A fixed random sample of 128 validation queries was processed
serially three times after 24 branch-covering warmups. All 384 top-1 outputs
matched the stored validation predictions. Timers include raw-spectrum preprocessing,
data-safe DreaMS, three SCCF spectrum projections, ranking the full supplied
pool against cached molecule projections, top-20 support checks, pair
reranking, coverage and feasibility-tree scoring, and the complete top-5
feasibility reranker only when routed. GPU stages are synchronized.
Frozen pair-reranker repeat-padding up to 512 rows is retained; this measures the
implemented numerical path, not an optimized batch-one kernel. Hardware is
one RTX~4080~SUPER, with one PyTorch CPU thread, two LightGBM threads and
zero data workers.

\begin{table}[htbp]
\centering\small
\caption{Validation latency in milliseconds per query. Passes reuse the same
128 queries and molecule projections. The fragment-proposal cache fills
on first access and is retained for repeats. The first part of pass~1
overlapped two CPU MCES workers; they were stopped before later passes.
Every observation is retained.}\label{tab:s7-validation-latency}
\begin{tabular}{@{}lrrrr@{}}
\toprule Pass / proposal-cache state & Mean & Median & P90 & P95 \\ \midrule
1, partly populated & 393.53 & 347.63 & 672.89 & 833.38 \\
2, repeated-query access & 33.07 & 36.10 & 39.60 & 40.64 \\
3, repeated-query access & 32.80 & 35.73 & 39.09 & 40.49 \\
\bottomrule
\end{tabular}
\end{table}

Initialization took 20.92~s; peak allocated GPU memory was 3,213.82~MiB.
Input-file reads, candidate retrieval, network/queueing, startup and unseen
molecule featurization/projection are outside the timer; cached molecule-table
gathers and CPU/GPU transfers are inside. Repeat observations are not 384
independent spectra. No matched MSAlign latency or overall efficiency advantage
is established.

\clearpage
\subsection{Secondary myopic-MCES distance at top~1}

This structural-distance analysis was added after the primary evaluation and
is outside the prespecified primary analysis. Nine rows use scores stored
during the primary evaluation. For six comparison rows, we recovered top-1 identities by scoring with
the frozen models on 19 October 2026, preserving the original numerical
settings and embeddings. Full validation candidate scores and original
test contributions reproduced bitwise; these additional predictions were
fixed before MCES analysis. Distances use uniform averaging over exact top-1
ties. Method seed
means average per-seed distances; they are not score ensembles. No query is
dropped. Larger errors can be represented by lower bounds: this is the
threshold-15 myopic metric with the stronger bound enabled, not an all-pair
exact-MCES calculation. Values are distances, not percentages, and lower is
better. Software versions are myopic-mces 1.3.2, PuLP 3.3.2, HiGHS 1.15.1 and
RDKit 2023.09.6. This distance must not be
confused with recall on an MCES-defined molecular split.

\begin{table}[htbp]
\centering\small\setlength{\tabcolsep}{5pt}
\caption{Myopic-MCES@1 over all 17,556 official-test queries. Balanced
values average within 2,997 connectivities first. All 15 prespecified rows are
complete: nine use stored scores from the primary evaluation and six use the additional
19 October 2026 frozen-model scoring. This secondary
rescoring changes no model, primary result or method selection.}
\label{tab:s7-posthoc-mces}
\begin{tabular}{@{}lrr@{}}
\toprule Method & Query-weighted MCES & Balanced MCES \\ \midrule
\RowSCCFMean & 17.9272 & 19.6247 \\
\RowMSAlignMean & 15.9703 & 17.3485 \\
\RowEmbCosMean & 20.1854 & 20.5638 \\
\RowEnsemble & 17.3173 & 18.9903 \\
\RowTopFivePair & 16.5784 & 18.2568 \\
\RowTopFivePairCov & 16.3510 & 17.9364 \\
\RowTopFiveCovDouble & 16.1300 & 17.6356 \\
\RowTopFiveScoreOnly & 15.7313 & 16.9967 \\
\RowTopFiveFeas & 15.0825 & 16.1591 \\
\RowTopTwentyPair & 16.6420 & 18.2116 \\
\RowTopTwentyPairCov & 16.2593 & 17.7240 \\
\RowTopTwentyCovDouble & 15.8571 & 17.1323 \\
\RowTopTwentyScoreOnly & 15.3641 & 16.3468 \\
\RowTopTwentyFeas & 14.7478 & 15.4508 \\
\RowFinal & 14.5603 & 15.1788
\\
\bottomrule
\end{tabular}
\end{table}

The nine original rows contain 36,345 distinct symmetric molecule pairs. The
six additional rows contain 8,606 pairs, sharing 8,139 with the original rows
and adding 467, for a union of 36,812. Every pair has a valid distance or
certified lower bound. We resolved 84 initially unsuccessful computations
using numerical certificates, including one threshold-15 cutoff certified by
a valid MIP dual bound and half-integer objective grid. One additional CBC
timeout was solved by HiGHS with zero MIP gap and primal/dual value 14.0.
No solver failure is replaced by an arbitrary penalty. The original distances
are unchanged.
These descriptive results change no method or primary contrast and
do not establish statistical significance or isolate a backbone effect.

\subsection{Additional ensemble and mass-prior contrasts}

The aggregation-matched ensembles use frozen per-seed scores.
Connectivity-balanced R@1 of the Stage~1 ensemble was 3.61 points
below the MSAlign ensemble (Stage~1 minus MSAlign: $-3.61$
[$-4.83$, $-2.44$]). The final system's R@20 was 4.85 points lower under
connectivity balancing [$-6.26$, $-3.44$]; the query-weighted contrast is
reported in the main text. In the precursor-mass analysis, the
final system's R@1 gain was 4.1 points [2.7, 5.6] smaller than the MSAlign
ensemble's. The fused point estimates were 37.88\% and 35.60\%, respectively,
but no interval was computed for their difference. These comparisons span
different encoders, metadata and training budgets and do not isolate the
molecular target.

\clearpage
\section{Stopped configurations and their scope}
\label{sec:supp-negative}

Table~\ref{tab:supp-negative} summarises negative or inconclusive results.
Historical formula-group splits and ordinary-DreaMS experiments provide
design evidence distinct from the official-split, data-safe protocol. Each
conclusion concerns the tested configuration rather than its entire model
family. The table covers complementary failure modes; it is neither an
exhaustive review nor a quantitatively pooled meta-analysis.

\footnotesize
\begin{longtable}{@{}>{\raggedright\arraybackslash}p{28mm}>{\raggedright\arraybackslash}p{45mm}>{\raggedright\arraybackslash}p{79mm}@{}}
\caption{Negative or inconclusive configurations, with the protocol and limit
of each conclusion. Official-split local-evidence methods and results are
described in Sections~S6.4 and S7.9.}\label{tab:supp-negative}\\
\toprule Configuration & Evaluated setting & Finding and scope \\ \midrule
\endfirsthead
\toprule Configuration & Evaluated setting & Finding and scope \\ \midrule
\endhead
Old-space recovery & Old-author weights transferred to canonical v1.5 molecular strings & Molecule-only, spectrum-only, low-learning-rate joint and distillation recipes did not restore retrieval. This is an alignment-migration failure, not evidence that released GLMR ignored spectra. \\
Frozen token adapter / naive last-block tuning & Historical formula-group split, ordinary DreaMS, matched Stage~1 baselines & Tested readouts and naive last-block adaptation did not clear matched retrieval gates. This does not bound a new objective or candidate-conditioned encoder. \\
Unweighted in-batch negatives & Historical 2k/2k formula-group pilot, mass cap256 & Adding valid batch ground truths at equal weight damaged early ranking despite false-negative masking. Weighting and curriculum were not established. \\
Fixed hard/random negatives & Historical 20k formula-group setup, mass cap256 & The fixed 50\% hard fraction reduced R@1/MRR relative to random negatives. No general rejection of hard-negative curricula follows. \\
True dual latent & Historical frozen DreaMS/ChemBERTa pilot; $M=4$, $N=8$, $d=512$, cap128 & Mean--max interaction remained below the global anchor despite Stage~1 KL. The anchor used cap256, so this is not a matched-cap method comparison. \\
Fragment-GINE & Historical formula-group split after full-training coverage expansion, ordinary DreaMS, natural top-5, bounded holdout & No R@1 increment over the same-budget continued Fragment-v2 control; the declared effect gate failed. The development-exposed pilot does not reject other graph/proposal supervision. \\
Learned peak--fragment alignment, fixed readout & Official/data-safe, semantics trained and assessed on training data, frozen top-5, official validation & Within-molecule semantic learning passed its prespecified criterion; the retrieval increment did not beat the matched coverage control. A gain beyond coverage was not established; local semantic learnability itself was not refuted. \\
Cross-candidate peak--fragment supervision & Official-train-only, connectivity-separated fit/calibration/evaluation and upstream OOF & Candidate supervision did not beat local-loss-only training or the coverage$\times$2 control on the isolated cohort. The experiment did not evaluate official validation or test and does not establish a benefit from scaling. \\
Fingerprint auxiliary / transfer & Historical ordinary-DreaMS, matched Stage~1 controls & Auxiliary fingerprint training and the tested frozen-backbone transfer did not establish a matched retrieval gain. This does not show fingerprint targets are universally unsuitable. \\
Structured feasibility head & Historical development-exposed formula-group validation, fixed strong score-fusion baseline & Extra route/shift/ppm information beat the collapsed control but missed the prespecified effect-size and connectivity-cluster uncertainty gate. Positive feature information was insufficient to establish the required retrieval gain. \\
MS/MS isotope satellites & Historical formula-group Stage~1 validation, observed fragment peak lists & Broad policies caused net damage; the narrow positive policy was unresolved. Molecular isotope verification requires linked MS1, not further tuning of fragment-list counts. \\
Property-matched shortcut control & Historical custom-pool matching; official-fold ordinary-DreaMS 20k counterfactual pilot & Tested matching/training recipes did not remove prior retention. A wider traceable candidate universe and full blind controls remain necessary for a clean claim. \\
\bottomrule
\end{longtable}
\normalsize

\section{Supplementary methods}\label{sec:si-methods}

The corresponding source, configurations and archived outputs are available
in the \href{https://github.com/ChristopherXMag/msms-retrieval-audit/releases/tag/v1.0.0}{version 1.0.0 reproducibility release}
(tag \texttt{v1.0.0};
\href{https://github.com/ChristopherXMag/msms-retrieval-audit/commit/e3524f7e62e63b4805d1a577d3076cba27f9ce28}{commit \texttt{e3524f7e62e6}}).
The repository maps reported findings to source and artifacts and documents
how to verify archived predictions and analyses. Original software and
author-trained model files use MIT; author-owned original predictions,
statistical summaries and figure source data use CC BY 4.0. Third-party terms
and upstream acquisition requirements are documented separately. Release
validation does not constitute a fresh end-to-end training or inference run.

\subsection{Historical formula-group split and representation screen}
\label{sec:si-historical}

Design evidence that predates the official protocol comes from a
formula-group development split of MassSpecGym v1.5 with
184,847/23,134/23,123 train/validation/test spectra, in which each molecular
formula occurs in one partition only. That work used the ordinary MoNA-tuned
DreaMS checkpoint and Morgan radius~3 inside the four-block composite. A
representation screen sampled 20,000 training spectra, trained each candidate
molecule target for five epochs and selected checkpoints lexicographically by
validation MRR, R@1 and R@20. The screen covered frozen ChemBERTa
\cite{ahmadChemBERTa2ChemicalFoundation2022,praski2026benchmarking}, binary
Morgan, smaller count composites, chirality and molecule-head controls. Its
results explain design choices; they are not official benchmark results and
are never pooled with official-split rows.

The coverage experiment on this split compared nested training subsets with
narrow or diversity-maximising molecule selection at matched row counts and
approximately matched optimisation steps. A full-coverage arm then expanded
the positive training support from 8,671 molecules in the 20,000-row screen
to all 22,934 training molecules, with molecule-balanced and all-row sampling
over three seeds; all-row sampling uses all 184,847 training spectra and
therefore changes both repeat exposure and the number of optimisation steps.

On the official split, the radius comparison holds the spectrum cache and
training protocol fixed while varying only the Morgan radius. A shallow
complementarity analysis compares SCCF with a Morgan-r2-only target over three
matched seeds and uses single-block Atom Pair, path and MACCS models trained
with the first prespecified seed only (chosen for computation, not by
validation score) to describe correctness-set overlap. Per-query differences
are averaged over matched seeds before bootstrap resampling. The single-block
oracle is descriptive and was not used to select a model.

\subsection{Comparison-method configurations}
\label{sec:si-external}

Table~\ref{tab:si-protocol-comparison} separates the protocols used for our
matched comparisons from selected related approaches. A mass-selected
candidate list and formula-free query processing are different conditions:
query formula information can enter peak annotations even when candidates
are selected only by mass. Formula-aware models may instead use a separately
validated formula predictor; such a pipeline must include that component's
provenance and prediction error \cite{liu2026massspecgymwild}.

\begingroup
\footnotesize
\setlength{\tabcolsep}{3pt}
\renewcommand{\arraystretch}{1.12}
\begin{table}[p]
\footnotesize
\caption{Input and evaluation scope of the current comparisons and selected
related methods. GT: ground truth; CE: collision energy. ``Adapted'' means
trained and evaluated here under our official v1.5 contract; it does not mean
numerical reproduction of a paper's original protocol or headline score.
No cross-study performance ranking is implied.}
\label{tab:si-protocol-comparison}
\centering
\begin{tabular}{@{}>{\raggedright\arraybackslash}p{23mm}>{\raggedright\arraybackslash}p{42mm}>{\raggedright\arraybackslash}p{22mm}>{\raggedright\arraybackslash}p{41mm}>{\raggedright\arraybackslash}p{21mm}@{}}
\toprule
Method & Query and candidate inputs & Query GT formula & Split / candidate protocol & Evaluated here? \\
\midrule
SCCF / final system & Data-safe DreaMS and SCCF; Stage~2 uses query peaks,
precursor/adduct and candidate fragments & No & Official v1.5
train/validation/test; mass candidates, cap~256 & Yes; proposed system and matched controls \\
MSAlign \cite{krzakala2026msalign} & Data-safe DreaMS, frozen MolDeBERTa,
CE and adduct encodings & No in our mass track & Same official v1.5
partitions/pools; released single-model MCES configuration & Yes; adapted and trained \\
EmbCos \cite{krzakala2026msalign} & Binned MS/MS and binary Morgan
fingerprints; no explicit metadata & No & Same official v1.5
partitions/pools; released EmbCos recipe & Yes; adapted and trained \\
MIST, MassSpecGym audit \cite{goldmanAnnotatingMetabolitemass2023,liu2026massspecgymwild}
& Formula-annotated peaks and fingerprint prediction & Oracle or predicted
formula, depending on variant & Published MassSpecGym mass/formula
protocols; formula source must be stated & No matched rerun \\
MVP, primary FormSpec \cite{chenLearningAllViews2026} & Formula-annotated
single/consensus spectra; molecular graphs and fingerprints & Yes for
reported FormSpec peak annotation & Published MassSpecGym MCES split;
mass/formula candidate tracks & No matched rerun \\
FLARE, reported benchmark \cite{chenFLAREFinegrainedLearning2026}
& Subformula-annotated peaks, known parent formula/adduct and molecular
node embeddings & Known parent formula in reported preprocessing & Published
MassSpecGym MCES split; mass/formula candidate tracks & No matched rerun \\
GLMR \cite{zhang2025glmr} & Spectrum/SMILES alignment, retrieved candidates
and context-conditioned molecule generation & Not for mass library;
used for formula library & Original MassSpecGym partitions/libraries;
MassRET-20k transfer & Historical diagnostics only; no matched v1.5 row \\
ICEBERG / GLACIER \cite{goldmanGeneratingMolecularFragmentation2024,wang2026glacier}
& Candidate molecular graphs and simulated spectra; released conditioning
includes adduct, CE and instrument & No query GT formula in the cited
mass challenge & Author-reported MassSpecGym mass challenge in Table~S1;
checkpoint/cohort not matched here & No matched rerun \\
\bottomrule
\end{tabular}
\par\medskip
\raggedright
The MSAlign source publication also reports formula-group splits and MCES
variants with validation/test roles exchanged; our adapted rows use one fixed
official partition assignment. Its seven-scorer late fusion and oracle-formula
filter are not reproduced as comparison rows. The separate precursor-mass
analysis in this study does not recreate that complete fusion system.
MVP also reports binned-spectrum variants; its FormSpec row above refers to
the formula-annotated representation, not every possible MVP configuration.
For FLARE, the table describes the parent-formula assumption in the reported
input representation, separately from whether the candidate library is
restricted by formula. No performance advantage is inferred from these
protocol differences. The score-only, coverage and feature-channel controls
within our system provide the primary attribution evidence.
\end{table}
\endgroup

Both external methods were adapted to the official train/validation
partitions, canonical-isomeric identities and original mass-cap~256 candidate
pools, preserving candidate counts and order, ground-truth coverage and the
held-out-connectivity negative exclusion used for SCCF.
For both adaptations, the released source was fixed to the
\href{https://github.com/ChristopherXMag/MSAlign-NeurIPS2026/tree/c2ef425b68749052a17b9cdd5f5f11709322654e}{MSAlign repository, commit \texttt{c2ef425b6874}}.

\emph{MSAlign.} We adapted the released single-model MSAlign v2 MCES
configuration \cite{krzakala2026msalign}. It uses the data-safe DreaMS
checkpoint with its released contrastive-head output, rather than the
pre-head state used by SCCF, and frozen MolDeBERTa-base-123M-mtr first-token
vectors (768 dimensions, at most 128 tokens; revision
\texttt{7d06a5f39098837a6f33ed095e2fa7706e30f8b9}). Features were cached for
6,491,031 molecule rows referenced by train/validation; 22,845 rows exceeded
the token limit. Identical effective token inputs share one float32 vector
while every candidate identity is retained. The shared alignment dimension is
512, hidden width 2,048 and dropout 0.30; the molecule branch has three hidden
blocks. The spectrum branch has a linear pre-projection, concatenated
64-dimensional collision-energy and adduct encodings and one hidden
post-projection. The adduct vocabulary is learned from training data only;
missing collision energy uses the released learned missing-value vector. The
trainable temperature starts at 0.07. AdamW uses learning rate $10^{-4}$,
weight decay $10^{-6}$, 2,000 warm-up steps, cosine decay and gradient
clipping at 5. Each seed uses batch 128, the ground truth plus 127 uniformly
sampled eligible candidates and 20,000 optimiser steps. Both foundation
encoders remain frozen. Gaussian precursor-mass scores and multi-model score
fusion from the original publication are not part of this single-model row.

\emph{EmbCos.} The EmbCos baseline distributed with the MSAlign code
\cite{krzakala2026msalign} uses the observed 100 most intense peaks, 0.1-Da
bins up to 1,005~Da and the released logarithmic intensity transform, without
a synthetic precursor peak or explicit metadata. Molecular inputs are binary
non-chiral Morgan-r2 fingerprints with 4,096 bits. The released
spectrum/molecule projections use three hidden 1,024-unit blocks and a linear
molecule branch, shared dimension 256 and dropout 0.25. AdamW uses learning
rate $10^{-4}$, weight decay $5\times10^{-5}$, 2,000-step warm-up with cosine
decay and clipping at 5; each seed uses batch 64, the ground truth plus 31
uniform candidates and 16,000 steps. The split-specific MGF source and its
six-decimal export rounding are retained.

Both methods use the same three training seeds as SCCF and select, per seed,
the checkpoint with maximal query-weighted official-validation R@1 among complete epochs and
the final partial epoch, keeping the earliest step on an exact tie. Epoch~0 is
ineligible; there is no early stopping, seed exclusion or further parameter
search. Paired validation intervals compare seed-mean query contributions with
matched SCCF and condition on the validation-selected checkpoints.

\subsection{Out-of-fold training data for the feasibility trees}
\label{sec:si-oof}

One hash-selected spectrum per connectivity was drawn from official training
data: 8,000 fit and 1,024 outer-calibration connectivities, fixed before any
rank or support information was known. The fit set is divided into two inner
folds (3,980 and 4,020 molecules). Three model families are trained: each
inner-fold family predicts the fold it excluded, and an endpoint family
predicts outer calibration. For each family, the held-out and
outer-calibration ground-truth connectivities and every alias of them in the
candidate universe are excluded from both the positive rows and the negative
identities of every upstream SCCF and pair model. A complete scan of the
7,220,331-row candidate universe found 18,249 aliases for 9,024 target
connectivities. Each family trains three fresh SCCF models with the same
prespecified training seeds for one
epoch on its complementary training rows (152,429, 153,282 and 187,123 positive
rows) and fits its own pair reranker with the recipe of the main text. Outer
calibration excludes all connectivities of the earlier training-cohort pilot.
The isolation is conditional on the released DreaMS features: encoder
pretraining was not repeated per fold.

The top-20 configuration reuses the nine family-specific SCCF fits, retrains
the pair rerankers and trees, and recomputes all shortlist statistics over the
actual 20-candidate shortlist. Each upstream checkpoint stores one fp16 vector
per unique candidate row with int32 query indices; under-filled training
pools use true-length masks without deleting queries or inserting candidates.

\subsection{Learned peak--fragment alignment, proposal audit and pilots}
\label{sec:si-local}

The learned peak--fragment reranker uses frozen data-safe DreaMS peak tokens
and an edge-aware GIN fragment encoder. Its local loss retains every
MAGMa-positive peak--fragment mask, masks unmatched peaks rather than treating
them as negatives, and contrasts annotated fragments within each training
molecule. At inference, deterministic atom-mask proposals (at most 64 per
candidate) admit only direct-fragment or precursor-neutral-loss matches within
20~ppm, with fixed hydrogen shifts. A null-normalised support, a
candidate-relative residual and intensity weighting produce a local score,
which the frozen readout adds as $0.25$ times its within-top-5 standardised
value to the pair-reranker score. The matched coverage-only control replaces
every feasible peak--proposal similarity by one while keeping observed peak
masses and intensities. The scorer preserves top-5 membership and falls back
for unsupported queries. All choices were fixed before that evaluation; the
design draws on MAGMa and learned fragmentation graphs
\cite{ridderAutomaticCompoundAnnotation2014,
goldmanGeneratingMolecularFragmentation2024,chenFLAREFinegrainedLearning2026}.

Because a local scorer can use only peaks that some candidate proposal can
explain, the proposal family was audited before any scorer was trained. On a
deterministic 20,000-spectrum sample of official training data, MAGMa
annotated the ground truth against the 100 most intense peaks at 20~ppm, with
fragmentation depth~3, at most six broken bonds and every alternative atom mask
retained \cite{ridderAutomaticCompoundAnnotation2014}. The audited family
contains connected components obtained by one non-ring single-bond cut or by
two cuts within the same ring, deterministically subsampled to a fixed cap. A
MAGMa-assigned peak counts as recovered only when one of its annotated atom
masks equals a proposal exactly. Recall is reported over MAGMa-assigned peaks
and intensity, overall and by heavy-atom and ring count.

A training-cohort pilot used only official-training molecules: 2,038 fit,
509 calibration and 2,038 evaluation connectivities, one spectrum per
molecule. Fresh single-seed SCCF models excluded held-out connectivities both
as positive training rows and as negative candidates; two inner folds provided
training lists, and outer calibration/evaluation connectivities were excluded
from the whole learned upstream pipeline. Matched local encoders used the
local loss alone or candidate listwise cross-entropy plus 0.1 times the local
loss; a doubled-coverage comparator tested score amplification. A subsequent
CPU LightGBM pilot on the same cohort learned residuals above the fixed
SCCF+coverage score from 10 score-derived features, the 36 features of the
main text, or 74 features including candidate descriptors, with three scorer
seeds and 25/50/100/200 trees chosen on calibration under the damage rule.
These pilots are development evidence only.

\subsection{Descriptive local molecular--spectral geometry}
\label{sec:si-geometry}

Motivated by the task dependence of molecular representation geometry
\cite{rottach2025topology}, a CPU-only diagnostic used existing
official-training spectra. The cohort contains 2,365 molecular-pair--condition
instances, 926 distinct canonical-isomeric molecule pairs and 355 formula
groups. Endpoints have different connectivities and share formula, adduct,
instrument type, coarse dataset cohort and exact known numeric collision
energy. Formula restricts this diagnostic cohort only; it is not a retrieval
input or filter. Missing conditions were excluded, and fine acquisition
provenance and collision-energy units could not be verified.

Each endpoint contributes two distinct processed peak lists, giving two
cross-molecule spectral views; these are repeat proxies, not independently
verified acquisitions. We retain the 100 most intense positive peaks below
precursor $m/z-1.5$~Da, require at least five peaks, apply square-root
intensity and $L_2$ normalisation, and use greedy cosine matching at 0.02~Da
without precursor shifts. With within-molecule repeat cosines $u_a,u_b$ and
$W_{ab}=(u_a+u_b)/2$, a large spectral difference requires both cross-view
cosines to be at least 0.05 below $W_{ab}$; a spectral cliff is such a
difference among pairs that a representation places close together.
Continuous spectral distance is one minus the mean of the four cross-molecule
cosines.

We compare chiral log-count Morgan-r2 cosine, raw block-normalised SCCF,
frozen first-token MolDeBERTa and the three trained MSAlign molecule-tower
outputs. Pairs have equal total weight across conditions. We report the cliff
fraction in the closest 10\% of pair mass and the normalised integral of the
prefix cliff-fraction curve over 0--25\%. A second analysis retains 1,849
anchored triple--condition instances from 81 formulas, scoring ordering
agreement with half credit for exact distance ties; weighted Spearman
correlation is auxiliary. Raw SCCF minus Morgan is the primary contrast, with
5,000 paired whole-formula bootstrap resamples and no multiplicity correction.
No model was fitted and no official validation or test array was read.

\subsection{Random seeds and evaluation details}
\label{sec:si-test-execution}

\paragraph{Random seeds.}
The ordered training seeds for SCCF, MSAlign and EmbCos are
20260621--20260623. Other seeds are 20260820--20260822 for matched
wrong-spectrum assignments; 20260926 for primary identity-key shortlist ties,
with 20260927 and 20260928 for sensitivity checks; 20261017 for the primary-test
bootstrap; 20261018 for latency sampling; and 20261112 for the validation
illustrations.

Data-safe DreaMS inputs for both spectrum readouts were generated with the
pinned checkpoint. The training-mean control uses the embedding mean over all 194,119
official-training spectra and their mean raw-peak spectrum, accumulated in
0.01-Da bins after per-spectrum normalisation, keeping the 100 highest bins
and normalising again; the two constants are separate training means, not an
embedding reconstruction. Matched wrong-spectrum donors share adduct and
parent mass rounded to six decimal places, differ in recomputed connectivity
and form three fixed closed one-to-one derangements; no
matching rule is relaxed.

Shortlist execution uses the primary identity-key seed. Exact-tie metrics average
uniformly within the reranked prefix and separately within the retained
Stage~1 tail; the two additional seeds are sensitivity checks only.
Boundary ties can therefore produce a small R@20 difference from standalone
Stage~1's full-pool uniform expectation. Fractional tied hits give sharp rescue/damage bounds
rather than an assumed joint distribution of two methods' tie outcomes.

Before test inference, numerical reproducibility was verified on all 19,429
validation queries: all 11 Stage~1 arrays reproduced bitwise.

The secondary MCES analysis was added after the primary test evaluation. Nine rows use
scores stored during the primary evaluation. Six comparator rows had retained ground-truth ranks
without full top-1 identities; an additional frozen-model scoring
pass recovered them after their complete validation scores reproduced bitwise,
and it reproduced every original test contribution. Myopic MCES@1 uses
threshold 15 and the stronger-bound option, preserving exact-tie expectations
and per-seed aggregation.

\subsection{Resource accounting}
\label{sec:si-resources}

Resource rows reuse recorded runner and process wall times, parameter counts
and logical array sizes. Runner times include each method's validation work;
process times also include start-up and input verification. Molecule-feature
storage is reported with its row universe, precision and compression.
Out-of-fold fitting is training-only cost and is separated from the models
deployed at inference. Shared frozen encoders, additional molecule encoders,
projection caches and candidate-specific evidence are listed separately;
upstream pretraining cost and missing extraction timings remain unknown. A
validation-only study times the frozen complete system on 128 fixed
random queries from raw spectrum to ranking (Section~S4).

\section{Supplementary results}\label{sec:si-results}

Values are percentages (MRR multiplied by 100) and differences are
percentage points with paired 95\% connectivity-cluster bootstrap intervals
unless stated otherwise. Official validation was used sequentially during
development; its results are development estimates.

\subsection{Validation development of the final system}\label{sec:si-val-system}

\begin{table}[h]
\caption{Official-validation method means and system configurations.
Single-model rows average three seeds' metrics; the remaining rows rerank the
Stage~1 ensemble shortlist. Bal.~R@1 is connectivity-balanced.}
\label{tab:si-val-system}
\centering\small
\begin{tabular}{@{}lrrrrr@{}}
\toprule
Configuration & R@1 & R@5 & R@20 & MRR & Bal.~R@1 \\
\midrule
SCCF, single model (seed mean) & 20.81 & 45.39 & 71.22 & 32.78 & 13.79 \\
MSAlign, single model (seed mean) & 25.50 & 51.64 & 77.69 & 38.09 & 19.16 \\
EmbCos, single model (seed mean) & 13.35 & 30.11 & 52.62 & 22.38 & 12.91 \\
\midrule
\RowEnsemble & 22.84 & 48.56 & 75.38 & 35.20 & 15.62 \\
\RowTopFivePair & 23.69 & 48.56 & 75.38 & 35.82 & 17.42 \\
\RowTopFivePairCov & 24.71 & 48.56 & 75.38 & 36.71 & 18.69 \\
\RowTopFiveFeas & 29.11 & 48.56 & 75.38 & 39.67 & 25.27 \\
\RowTopTwentyFeas & 30.62 & 57.50 & 75.38 & 43.00 & 29.06
\\
\RowFinal & 31.43 & 57.28 & 75.38 & 43.48 & 29.81 \\
\bottomrule
\end{tabular}
\end{table}

\emph{Top-5 feasibility reranker.} With out-of-fold upstream training lists,
the fixed 50-tree model reached 29.11\% R@1 and 39.67\% MRR on validation. It
exceeded the pair reranker by 5.42 R@1 points and pair + coverage by 4.40
points [3.56, 5.31]. The matched score-only tree and coverage$\times$2 control
reached 26.71\% and 25.72\%. Its 15,020 supported queries were 77.3\% of
validation; unsupported queries keep the anchor order. Follow-up diagnoses
located much of the gain in queries where the ground truth had stronger
feasibility evidence than its score-selected challenger. Among 531
same-formula queries with equal coverage, the feasibility tree minus the
score-only tree was $-0.38$ points [$-2.13$, 1.38]; in a secondary
same-formula, equal-positive-count slice (647 queries) the
connectivity-balanced contrast was $-1.88$ [$-3.85$, $-0.22$] (exploratory).
Four fixed feature-group ablations gave no improvement over the full model.

\emph{Top-20 feasibility reranker.} The out-of-fold outer calibration passed
its continuation rule (R@1 60.94\% to 61.23\%, interval of the gain
[$-0.98$, 1.56]). On validation the top-20 reranker reached 30.62/57.50/75.38\%
R@1/R@5/R@20 and 43.00\% MRR, gaining 1.52 R@1 points [0.78, 2.23], 8.94 R@5
points [7.43, 10.48] and 3.33 MRR points [2.69, 4.01] over the top-5 reranker;
connectivity-balanced R@1 rose from 25.27\% to 29.06\% (+3.79 [3.05, 4.55]).
It rescued 767 queries and damaged 472, a damage rate of 8.35\% of the 5,655
queries the top-5 reranker ranked correctly, above the prespecified 8\% limit;
each tree seed also exceeded it. The gain closes 3.3\% of the top-5 reranker's
R@20$-$R@1 gap. Rank accounting found 563 newly correct ground truths from
original ranks 6--20 but 268 net losses among ranks 1--5. Tree support fell
from 15,020 to 12,544 queries, and 190 of the 472 damages occurred among the
2,476 queries that lost support.

\begin{table}[h]
\caption{Matched top-20 controls on official validation. All rows share the
natural top-20 shortlist and R@20 = 75.38\%.}
\label{tab:si-top20-controls}
\centering\small
\begin{tabular}{@{}lrrr@{}}
\toprule Configuration & R@1 & R@5 & MRR \\ \midrule
\RowTopTwentyPair & 23.66 & 49.33 & 36.03 \\
\RowTopTwentyPairCov & 24.95 & 51.95 & 37.66 \\
\RowTopTwentyCovDouble & 26.18 & 54.29 & 39.25 \\
\RowTopTwentyScoreOnly & 28.03 & 55.55 & 40.72 \\
\RowTopTwentyFeas & 30.62 & 57.50 & 43.00 \\
\bottomrule
\end{tabular}
\end{table}

The top-20 reranker exceeded its score-only tree by 2.59 R@1 points
[2.13, 3.08] and the coverage$\times$2 control by 4.45 points [3.85, 5.09]
(Table~\ref{tab:si-top20-controls}).

\emph{Support-gated fallback.} We retained the top-20 configuration for its
R@1 and MRR gains by author judgement despite the failed gate, and specified
the fallback after inspecting validation support strata. No model was
refitted and no threshold was searched; the revised system was frozen before
test evaluation. Replaying the fixed fallback rule on cached
branch outputs changed the 2,476 affected queries and reached
31.43/57.28/75.38\% R@1/R@5/R@20 and 43.48\% MRR. Relative to the top-20
reranker, R@1 improved 0.80 points [0.45, 1.25] and MRR 0.47 [0.26, 0.75],
while R@5 fell 0.22 points (connectivity-balanced $-0.38$ [$-0.63$, $-0.13$]).
Against the common top-5 baseline, damages fell from 472 to 282 (4.99\%) and
rescues from 767 to 733. These intervals condition on fixed fits and on a rule
motivated by validation outcomes.

\emph{Feature-channel control.} With real pair and anchor scores fixed and
donor spectra supplied only to the spectrum-derived tree features, all three
donor assignments gave positive real-minus-wrong R@1 intervals (gains
3.29--3.43 points) and positive MRR intervals. The one-sided 95\% upper bounds
of the wrong-minus-anchor gain were 0.68--0.85 R@1 points, above the fixed
0.25-point equivalence margin, so strict attribution of the gain to the
channel remains inconclusive.

\subsection{Stage~1 and comparison methods on official validation}
\label{sec:si-val-stage1}

\begin{table}[h]
\caption{Radius-2 SCCF on 19,429 official-validation spectra. Seed columns are
single models; the mean column gives mean $\pm$ sample standard deviation.
Train-mean and zero controls replace only the query spectrum representation.}
\label{tab:si-stage1-val}
\centering\scriptsize
\setlength{\tabcolsep}{2pt}
\begin{tabular}{@{}lrrrrrr@{}}
\toprule
Metric & Seed 21 & Seed 22 & Seed 23 & Mean & Train mean & Zero \\
\midrule
R@1 & 21.16 & 20.49 & 20.78 & 20.81 $\pm$ 0.34 & 13.88 $\pm$ 0.13 & 5.96 $\pm$ 3.45 \\
R@5 & 46.50 & 45.55 & 44.12 & 45.39 $\pm$ 1.20 & 31.16 $\pm$ 0.61 & 17.10 $\pm$ 8.22 \\
R@20 & 72.11 & 71.92 & 69.62 & 71.22 $\pm$ 1.39 & 51.99 $\pm$ 0.55 & 37.17 $\pm$ 8.45 \\
MRR & 33.29 & 32.67 & 32.39 & 32.78 $\pm$ 0.46 & 22.63 $\pm$ 0.27 & 12.92 $\pm$ 5.10 \\
Bal.~R@1 & 13.76 & 13.64 & 13.95 & 13.79 $\pm$ 0.15 & 4.96 $\pm$ 0.14 & 3.16 $\pm$ 0.94 \\
\bottomrule
\end{tabular}
\end{table}

Real spectra exceeded the training-mean control by 7.12, 6.66 and 7.00 R@1
points for the three checkpoints, with intervals [5.47, 8.89], [5.01, 8.70] and
[5.54, 8.54] (Table~\ref{tab:si-stage1-val}). A single-seed optimisation control
that trained SCCF for five epochs with MSAlign-style projection and metadata
heads reduced training cross-entropy from 1.78 to 0.16 but reached only
19.94\% validation R@1, below the one-epoch baseline (21.16\%).

MSAlign seeds completed 20,000 optimiser steps (selected steps 16,687, 19,721
and 20,000) and EmbCos seeds 16,000. MSAlign reached
$25.50\pm0.14$/$51.64\pm0.45$/$77.69\pm0.18$/$38.09\pm0.13$\% R@1/R@5/R@20/MRR
and EmbCos $13.35\pm0.69$/$30.11\pm0.24$/$52.62\pm0.33$/$22.38\pm0.46$\%.
MSAlign exceeded SCCF by 4.69 R@1 points [2.27, 6.80] (balanced 5.37
[4.40, 6.34]); EmbCos was 7.46 points lower [$-9.86$, $-5.16$] (balanced $-0.88$
[$-1.81$, 0.06]). MSAlign's validation zero and training-mean R@1 were 9.32\%
and 17.09\%; EmbCos's were 6.20\% and 0.66\%. Across three seeds, SCCF minus a
Morgan-r2-only target changed query-weighted R@1/R@5/R@20/MRR by
+2.69/+4.06/+1.93/+2.74 points; after connectivity balancing, R@1 changed by
$-0.18$ [$-0.96$, 0.60] while R@5 and R@20 gained 1.31 [0.33, 2.29] and 3.30
[2.28, 4.31]. Single-block models (first seed) had pairwise Jaccard overlaps of
0.24--0.36 between their rank-1-correct sets, and an oracle over the four
blocks reached 35.08\% R@1; this oracle is descriptive and mixes
representation with optimisation noise.

\subsection{Historical design screen and coverage experiments}
\label{sec:si-historical-results}

The historical screen changed only the molecule target on the formula-group
split with frozen ordinary DreaMS, mass-cap 256 candidates, candidate
cross-entropy, a fixed budget and one seed (Table~\ref{tab:si-target-screen}).
Explicit fingerprints were substantially stronger than frozen ChemBERTa
targets, and count and multi-view targets improved on binary Morgan. The
four-block composite (radius 3) improved R@1/R@20/MRR over the preceding
three-block composite by 2.47/1.76/2.75 points, all intervals excluding zero.
A non-chiral Morgan-r3 control was 3.50 points worse than its chiral
counterpart, and a learned block-gated head 3.93 points worse than the shared
MLP.

\begin{table}[h]
\caption{Selected molecule targets in the historical formula-group screen
(20,000 training spectra, single seed, custom test partition). Design
evidence, not official benchmark results.}
\label{tab:si-target-screen}
\centering\small
\begin{tabular}{@{}lrrrr@{}}
\toprule
Molecule target & R@1 & R@5 & R@20 & MRR \\
\midrule
ChemBERTa-MLM mean & 23.18 & 48.14 & 69.80 & 35.09 \\
ChemBERTa-MTR mean & 25.36 & 50.72 & 71.12 & 37.32 \\
Binary chiral Morgan r2 & 40.52 & 64.89 & 81.53 & 52.04 \\
Count Morgan r2 + atom pair & 41.04 & 68.58 & 83.53 & 53.77 \\
Morgan r2 + path + MACCS & 42.65 & 67.62 & 82.56 & 53.96 \\
Four-block composite, r3 & 45.12 & 70.23 & 84.32 & 56.71 \\
\bottomrule
\end{tabular}
\end{table}

On that split's validation partition (23,134 spectra, 2,974 connectivities),
rank-1 correctness among 965 molecules with at least four spectra had ICC(1)
0.756; 44.9\% were wrong and 28.4\% right for every spectrum. A
within-molecule oracle raised R@1 from 46.46\% to 59.65\%. Peak count and
precursor mass hardly differed between these groups, whereas training
proximity correlated with molecule-level R@1 (Spearman 0.550). In row-count-
and step-matched coverage experiments, the diversity-maximising training
subset beat the narrow subset in all nine comparisons (three sizes, three
permutations), by 6.1 R@1 points on average. Expanding the positive training
support from 8,671 to all 22,934 molecules improved three-seed R@1/MRR/R@20 by
3.79/3.41/3.31 points with molecule-balanced sampling
(Table~\ref{tab:si-coverage}).

\begin{table}[h]
\caption{Three-seed coverage experiment on the historical validation
partition (query-weighted).}
\label{tab:si-coverage}
\centering\small
\begin{tabular}{@{}lrrrr@{}}
\toprule
Training arm & R@1 & R@5 & R@20 & MRR \\
\midrule
Original 20,000-spectrum control & 45.58 & 70.80 & 85.49 & 57.19 \\
Full coverage, molecule-balanced & 49.37 & 74.62 & 88.81 & 60.61 \\
Full coverage, all rows & 51.62 & 74.92 & 88.30 & 62.28 \\
\bottomrule
\end{tabular}
\end{table}

\subsection{Matched wrong spectra and near-isomers}\label{sec:si-wrong}

On the 2,155-query validation slice with exact parent mass, same adduct and
different connectivity, the Stage~1 checkpoints averaged 14.01\% R@1 with real
spectra and 12.08\% with wrong spectra; seven of nine R@1 and all nine MRR
real-minus-wrong intervals excluded zero (balanced 15.06\% versus 10.89\%).
Training proximity correlated with connectivity-level wrong-spectrum R@1
(Spearman 0.285 [0.207, 0.360]), also after controlling for donor--target
similarity and candidate count (0.154 [0.056, 0.243]). Donor--target
similarity was the more direct marker (Table~\ref{tab:si-wrong}): from its
lowest to its highest fifth, wrong-spectrum R@1 rose by 22.2 points
[13.9, 30.9] while the real-minus-wrong gap shrank by 10.8 points
[$-17.7$, $-4.6$]. A matched wrong spectrum is therefore not information-null
when donor and recipient are close isomers. Table~\ref{tab:si-val-controls}
gives the corresponding validation control table for all methods.

\begin{table}[h]
\caption{Connectivity-balanced validation R@1 across extreme fifths of
training proximity and donor--target SCCF similarity, averaged over two
checkpoints and three assignments.}
\label{tab:si-wrong}
\centering\scriptsize
\begin{tabular}{@{}llrrrrr@{}}
\toprule
Stratifier & Fifth & Groups & Spectra & Real & Wrong & Real $-$ wrong \\
\midrule
Training proximity & Lowest & 78 & 307 & 5.56 & 0.29 & 5.27 \\
Training proximity & Highest & 78 & 1005 & 18.39 & 16.41 & 1.98 \\
Donor--target similarity & Lowest & 77 & 231 & 10.06 & 0.79 & 9.27 \\
Donor--target similarity & Highest & 79 & 904 & 21.42 & 23.00 & $-1.58$ \\
\bottomrule
\end{tabular}
\end{table}

\begin{table}[h]
\caption{Validation spectrum-replacement controls on the matched slice (2,155
queries, 388 connectivities), query-weighted R@1; uplift ranges over three
assignments, $^\ast$ all intervals exclude zero. Candidate-only 5.28\%;
random 0.39\%.}
\label{tab:si-val-controls}
\centering\scriptsize
\setlength{\tabcolsep}{3pt}
\begin{tabular}{@{}lrrrrrr@{}}
\toprule
Method & Real & Wrong & Zero & Train mean & Query uplift & Balanced uplift \\
\midrule
SCCF (Stage~1 ensemble) & 15.08 & 12.36 & 3.20 & 7.29 & 2.65--2.78$^\ast$ & 5.44--5.68$^\ast$ \\
MSAlign ensemble & 17.31 & 13.63 & 7.75 & 11.28 & 3.62--3.71$^\ast$ & 7.44--7.74$^\ast$ \\
EmbCos ensemble & 9.15 & 6.15 & 1.86 & 0.19 & 2.93--3.09$^\ast$ & 5.89--5.99$^\ast$ \\
\RowFinal & 21.35 & 13.09 & 3.81 & 7.89 & 8.21--8.31$^\ast$ & 15.27--15.39$^\ast$ \\
\bottomrule
\end{tabular}
\end{table}

\subsection{Molecule-level clustering, proximity and the coverage intervention}
\label{sec:si-molecule}

On validation, the 852 connectivities with at least four spectra (70\% of
queries) gave ICC(1) 0.80 [0.76, 0.83] for the Stage~1 ensemble (all wrong
61.7\%, null 15.5\%), 0.72 [0.67, 0.75] for the final system (all wrong 48.2\%,
null 8.6\%), 0.70--0.74 for MSAlign seeds and 0.43--0.48 for EmbCos seeds; the
within-molecule oracle gains were 10.9 and 16.7 points for the Stage~1
ensemble and the final system. Table~\ref{tab:si-proximity} gives validation
proximity strata. The connectivity-level Spearman correlation between
proximity and Stage~1 R@20 was 0.44 [0.41, 0.47].

\begin{table}[h]
\caption{Official-validation results by training proximity of the query
molecule (query-weighted R@1 and R@20).}
\label{tab:si-proximity}
\centering\scriptsize
\begin{tabular}{@{}lrrrrr@{}}
\toprule
Proximity & Queries & Connectivities & Stage~1 R@1 & Stage~1 R@20 & Final R@1 \\
\midrule
$<0.60$ & 236 & 34 & 5.93 & 19.92 & 8.90 \\
$[0.60,0.70)$ & 3286 & 810 & 6.15 & 41.08 & 14.52 \\
$[0.70,0.80)$ & 6746 & 1510 & 14.04 & 70.92 & 27.35 \\
$[0.80,0.90)$ & 5012 & 636 & 36.21 & 92.12 & 45.19 \\
$\geq0.90$ & 4149 & 227 & 35.19 & 92.75 & 36.11 \\
\bottomrule
\end{tabular}
\end{table}

\emph{Coverage intervention.} The cohort A was drawn (seeded) as 200 of the
212 validation connectivities whose every ground-truth row had baseline
proximity at least 0.90. The targeted arm removed all 17,504 training spectra
of the 1,219 connectivities within cosine 0.80 of any A molecule; the random
arm removed the same number of spectra from connectivities below 0.80 to every
validation molecule (7,550 connectivities touched, 440 fully removed).
Negative candidate pools were unchanged. Before training, A's median
proximity dropped by 0.149 under targeted removal and was unchanged under
random removal; 2,691 connectivities (B-clean) kept every neighbour in both
arms and 293 (B-spill) lost some. Both arms used the first two prespecified
training seeds with the frozen Stage~1 protocol (Table~\ref{tab:si-r6}). In A, the
targeted minus random change was $-29.0$ R@20 points [$-41.6$, $-18.7$] and
$-24.6$ R@1 points [$-36.8$, $-15.3$]; in B-clean it was $+0.88$
[$-0.58$, 2.37] and $-0.23$ [$-1.45$, 0.98]. Within A and B-spill, larger
proximity drops were associated with larger R@20 losses (Spearman $-0.19$
[$-0.31$, $-0.05$] and $-0.40$ [$-0.50$, $-0.29$]). The four fits took about
20 minutes each on one RTX 4080 SUPER.

\begin{table}[h]
\caption{Coverage intervention on official validation, two-seed mean,
query-weighted. Full-train values reuse the existing seed-matched models.}
\label{tab:si-r6}
\centering\scriptsize
\begin{tabular}{@{}llrrr@{}}
\toprule
Cohort & Metric & Full train & Targeted removal & Random removal \\
\midrule
A (200) & R@20 & 93.14 & 62.04 & 91.02 \\
A (200) & R@1 & 31.53 & 4.94 & 29.55 \\
B-clean (2,691) & R@20 & 61.46 & 61.98 & 61.10 \\
B-clean (2,691) & R@1 & 13.12 & 14.00 & 14.23 \\
B-spill (293) & R@20 & 89.95 & 66.16 & 89.41 \\
B-spill (293) & R@1 & 39.09 & 10.15 & 37.21 \\
\bottomrule
\end{tabular}
\end{table}

\subsection{Aggregation-matched comparison on validation}\label{sec:si-r3}

With three-seed mean-logit ensembles for every method, validation R@1/R@20
were 22.84/75.38\% for SCCF, 27.20/79.47\% for MSAlign and 15.54/56.46\% for
EmbCos. The final system exceeded the MSAlign ensemble by 4.23 R@1 points
[1.80, 6.78] (balanced 9.05 [7.67, 10.47]) and was 4.09 R@20 points lower
[$-5.64$, $-2.61$].

\subsection{Precursor-mass prior on validation}\label{sec:si-r4}

Fusion weights were fitted by two-fold connectivity-disjoint cross-fitting on
validation, maximising candidate likelihood without R@1 selection; the test
rows of the main text use a single refit on all of validation. Cross-fitted
validation gains were 9.12 R@1 points [7.49, 10.86] for the Stage~1 ensemble
(22.84 to 31.96\%), 8.86 [7.41, 10.48] for the MSAlign ensemble, 8.47
[7.08, 10.02] for the EmbCos ensemble and 5.99 [4.83, 7.27] for the final
system (31.43 to 37.42\%). The mass prior alone reached 6.72\% R@1. Neutral
masses use float64 arithmetic from the published precursor $m/z$ and adduct,
and candidate masses are RDKit exact masses. For the final system the prior is
fused only within its executed shortlist, so shortlist membership and R@20 are
unchanged.

\subsection{Collision-energy availability}\label{sec:si-ce}

\begin{table}[h]
\caption{R@1 by collision-energy availability (query-weighted; balanced in
parentheses). Strata differ in composition; these are not causal estimates.}
\label{tab:si-ce}
\centering\scriptsize
\setlength{\tabcolsep}{3pt}
\begin{tabular}{@{}llrrrrr@{}}
\toprule
Split & Stratum & Queries / conn. & Stage~1 ens. & MSAlign & MSAlign ens. & Final \\
\midrule
Validation & CE present & 9,975 / 2,370 & 10.99 (11.20) & 14.76 (15.01) & 16.58 (16.69) & 24.55 (26.78) \\
Validation & CE missing & 9,454 / 1,080 & 35.35 (27.18) & 36.83 (28.67) & 38.41 (30.04) & 38.68 (36.99) \\
Test & CE present & 10,162 / 2,339 & 13.70 (12.45) & 15.30 (15.03) & 16.57 (16.35) & 26.52 (28.03) \\
Test & CE missing & 7,394 / 894 & 31.05 (25.39) & 35.66 (25.41) & 35.73 (26.90) & 37.29 (32.39) \\
\bottomrule
\end{tabular}
\end{table}

Spectra without a recorded collision energy came from fewer, more heavily
replicated molecules and were easier for every method
(Table~\ref{tab:si-ce}). The final system's margin over the MSAlign ensemble
was larger where collision energy was recorded.

\subsection{Local evidence: proposal coverage and development pilots}
\label{sec:si-local-results}

On 20,000 official-training spectra, MAGMa explained 53.82\% of the top-100
peaks and 77.06\% of their intensity. Of this MAGMa-assigned evidence, the
untruncated proposal family recovered an exact atom mask for 26.88\% of peaks
and 34.75\% of intensity (Table~\ref{tab:si-proposals}). Intensity recall was
40.90\% for molecules with at most 20 heavy atoms, 27.46\% above 40 heavy
atoms and 30.27\% with at least three rings. Using five top-ranked decoys per
query as a proxy, the ground truth exceeded the strongest decoy in
mass-feasible intensity coverage for 37.1--38.1\% of queries at caps 32--128.

\begin{table}[h]
\caption{Exact atom-mask recovery of MAGMa-assigned evidence by the proposal
family on 20,000 official-training spectra. Retention is relative to the
untruncated family.}
\label{tab:si-proposals}
\centering\small
\begin{tabular}{@{}lrrrr@{}}
\toprule
Cap & Mean proposals & Peak recall & Intensity recall & Retention \\
\midrule
32 & 31.40 & 15.89 & 24.25 & 69.79 \\
64 & 57.34 & 23.01 & 31.31 & 90.09 \\
128 & 78.44 & 26.47 & 34.40 & 98.99 \\
Untruncated & -- & 26.88 & 34.75 & 100.00 \\
\bottomrule
\end{tabular}
\end{table}

In a historical 1,000-spectrum audit on the formula-group split, 95--96\% of
the peaks explained by released GLACIER predictions \cite{wang2026glacier}
were already MAGMa-assigned, and adding GLACIER raised intensity coverage only
from 76.70\% to 78.29--78.83\%. The checkpoint's training-fold membership could
not be verified and 472 collision energies were imputed, so the audit is
descriptive.

The learned peak--fragment encoder passed its predeclared semantic gate on a
connectivity-disjoint training subset. Read out once on validation as an
additive term on the top-5 pair reranker (R@1 23.69\%), it added 0.93 points,
versus 1.02 for constant feasible-edge similarities and 0.49 for a
shuffled-label encoder; learned and constant rankings were identical for
99.1\% of queries. On the 2,155 matched queries, real coverage added 0.88 R@1
points to the pair reranker, against 0.14--0.19 points under wrong donor
spectra, with positive real-minus-wrong intervals.

\begin{table}[h]
\caption{Development evidence on 2,038 official-training evaluation molecules
(one spectrum each) with a fresh single-seed out-of-fold SCCF retriever. All
rows share R@5 = 75.47\% and R@20 = 89.35\%.}
\label{tab:si-pilot}
\centering\small
\begin{tabular}{@{}lrr@{}}
\toprule Configuration & R@1 & MRR \\ \midrule
SCCF + coverage & 52.26 & 63.23 \\
Coverage$\times$2 & 53.83 & 64.25 \\
Coverage + local loss only & 53.93 & 64.28 \\
Coverage + listwise + 0.1 local loss & 53.78 & 64.23 \\
Tree, 10 score features & 54.96 & 64.92 \\
Tree, 36 features & 56.67 & 65.99 \\
Tree, 36 features + candidate descriptors & 56.92 & 66.23 \\ \bottomrule
\end{tabular}
\end{table}

In this cohort (Table~\ref{tab:si-pilot}), listwise candidate supervision minus
the local loss alone was $-0.15$ R@1 points [$-0.39$, 0.05], and the two agreed
on the top candidate for 98.7\% of queries. The 36-feature tree exceeded
coverage$\times$2 by 2.85 points [1.77, 3.93] and the score-only tree by 1.72
[0.79, 2.65]; candidate descriptors added 0.25 [$-0.59$, 1.08]. Among 248
same-formula, equal-coverage pairs the ground-truth win rate changed from
71.37\% to 71.77\% (interval crossing zero).

\subsection{Local molecular--spectral geometry}\label{sec:si-geometry-results}

The training-set geometry diagnostic did not establish an advantage of raw
SCCF over raw Morgan (Table~\ref{tab:si-geometry}). SCCF minus Morgan changed
the nearest-10\% cliff fraction by $+0.80$ points [$-8.07$, 6.31], the
normalised 0--25\% cliff area by $+2.01$ [$-1.19$, 5.18] and anchored ordering
agreement by $+2.40$ [$-0.44$, 5.52]; every interval includes zero. Raw
MolDeBERTa had fewer nearest-neighbour cliffs but lower ordering agreement, and
the trained MSAlign spaces had higher distance correlations and ordering
agreement without fewer cliffs.

\begin{table}[h]
\caption{Training-set geometry diagnostics (not retrieval accuracies). C10:
cliff percentage in the closest 10\% of pair mass; C25: mean prefix cliff
percentage over 0--25\%; Ord.: ordering agreement (\%); $\rho$: weighted
Spearman correlation. MSAlign rows are trained molecule towers.}
\label{tab:si-geometry}
\centering\small
\begin{tabular}{@{}lrrrr@{}}
\toprule
Space & C10 & C25 & Ord. & $\rho$ \\
\midrule
Raw Morgan-r2 & 63.68 & 67.19 & 67.94 & 0.4095 \\
Raw SCCF & 64.48 & 69.19 & 70.34 & 0.4267 \\
Raw MolDeBERTa & 55.45 & 58.01 & 65.10 & 0.4595 \\
MSAlign seed 20260621 & 58.57 & 59.80 & 75.97 & 0.5767 \\
MSAlign seed 20260622 & 58.00 & 59.69 & 75.63 & 0.5814 \\
MSAlign seed 20260623 & 56.75 & 58.91 & 75.29 & 0.5892
\\
\bottomrule
\end{tabular}
\end{table}

\subsection{Resource accounting}\label{sec:si-resources-results}

\begin{table}[h]
\caption{Recorded development components. Params: trainable alignment
parameters per model (millions). Time: sum of three runner wall times
including validation (minutes), excluding feature extraction. Storage:
molecular input arrays (GiB); SCCF and EmbCos cover 7.22M rows and MSAlign
6.49M referenced rows, with different precision and compression.}
\label{tab:si-resources}
\centering\small
\begin{tabular}{@{}lrrr@{}}
\toprule Method & Params & Time & Storage \\ \midrule
SCCF & 19.24 & 71.12 & 27.76 \\
MSAlign & 13.79 & 101.27 & 18.57 \\
EmbCos & 13.71 & 18.95 & 3.44
\\
\bottomrule
\end{tabular}
\end{table}

MSAlign additionally uses a frozen 88.23M-parameter MolDeBERTa encoder, whose
feature extraction took 85.84~min; DreaMS cache generation took 5.71~min for
SCCF's readout and 5.34~min for MSAlign's, and each train/validation spectrum
cache occupies 0.815~GiB. The SCCF sparse arrays occupy 27.76~GiB, 20.55~GiB of
which are path counts. The deployed top-20 branch uses three SCCF heads, three
pair rerankers (1.07M parameters each) and three 50-tree models; the top-5
branch adds three pair rerankers and three tree models. Training the trees
required nine auxiliary out-of-fold Stage~1 fits (156.78~min of runner time).
In batch-one validation timing on 128 fixed random queries, the median
raw-spectrum-to-ranking latency was 35.73~ms (95th percentile 40.49~ms) with
cached molecule projections and fragment proposals and 347.63~ms (833.38~ms)
when proposals were first computed; full timing conditions are in
Section~S4. Upstream pretraining cost is unknown, and overall efficiency
superiority over the comparison methods is not established.

\bibliographystyle{plain}
\bibliography{references}